\documentclass[
 reprint,
 showpacs, preprintnumbers,
 amsmath, amssymb,
 aps,
 pra,
 floatfix,
 superscriptaddress
]{revtex4-1}

\usepackage{txfonts}

\usepackage{graphicx}
\usepackage{tikz}
\usepackage{xcolor}
\usetikzlibrary{arrows}

\makeatletter
\newcommand{\colorcaption}[2][]{%
	\begingroup%
	\renewcommand{\@caption@fignum@sep}{. }%
	\caption[#1]{#2}%
	\endgroup%
}
\makeatother

\usepackage{dcolumn}
\usepackage{bm}
\usepackage{bbm}
\usepackage{upgreek}
\usepackage{hyperref}
\usepackage{blindtext}

\usepackage{changes}

\usepackage{dsfont}

\definecolor{darkgreen}{RGB}{0 100 0}

\usepackage{pdfpages}
\makeatletter
\AtBeginDocument{\let\LS@rot\@undefined}
\makeatother

\begin{document}

	\title{Skew Andreev  reflection in ferromagnet/superconductor junctions}
	
	\author{Andreas Costa}%
	\email[Corresponding author: ]{andreas.costa@physik.uni-regensburg.de}
 	\affiliation{Institute for Theoretical Physics, University of Regensburg, 93040 Regensburg, Germany}
    
    \author{Alex Matos-Abiague}
    \affiliation{Department of Physics and Astronomy, Wayne State University Detroit, Michigan 48201, USA}
 	
	\author{Jaroslav Fabian}%
 	\affiliation{Institute for Theoretical Physics, University of Regensburg, 93040 Regensburg, Germany}

	\date{\today}

    \begin{abstract}
        
        Andreev~reflection~(AR) in ferromagnet/superconductor junctions is an indispensable spectroscopic tool for measuring spin polarization. We study theoretically how the presence of a thin semiconducting interface in such junctions, inducing Rashba and Dresselhaus spin-orbit~coupling, modifies AR~processes. The interface gives rise to a momentum- and spin-dependent scattering potential,
        making the AR~probability strongly asymmetric with respect to the sign of the incident electrons' transverse momenta. This \emph{skew AR} creates spatial charge carrier imbalances and transverse Hall~currents in the ferromagnet. We show
        that the effect is giant, compared to the normal regime. 
        We provide a quantitative analysis and a qualitative picture of this phenomenon, and finally show that skew AR also leads to a widely tunable transverse supercurrent response in the superconductor.

    \end{abstract}

    \maketitle

        Due to the extraordinary properties occurring at their interfaces, ferromagnet/superconductor~(F/S) heterostructures attract considerable interest~\cite{Eschrig2011,Linder2015,Gingrich2016}. Such junctions might not only offer novel tools for controlling and measuring charge and spin~currents, but might also bring new functionalities into spintronics~devices.
    
    
        While early efforts focused on detecting spin-polarized quasiparticles in superconductors via spin transport experiments~\cite{Tedrow1971,Tedrow1973,Meservey1994}, current progress in the 
        rapidly growing field of superconducting spintronics~\cite{Linder2015} opened several promising perspectives, ranging from the observation of long spin~lifetimes and giant magnetoresistance~effects~\cite{Yang2010} to the generation and successful manipulation of superconducting spin~currents~\cite{Wakamura2015,Beckmann2016,Bergeret2016,Espedal2017,Linder2017,Ouassou2017,Jeon2018,Montiel2018}. But the interplay of magnetism and superconductivity gets even more interesting when spin-orbit coupling~(SOC) of the Rashba~\cite{Bychkov1984} and/or Dresselhaus~\cite{Dresselhaus1955}~type is present~\cite{Fabian2004,Fabian2007}. Prominent examples are spin-triplet pairing mechanisms~\cite{Bergeret2001,Volkov2003,Keizer2006,Halterman2007,Eschrig2008,Eschrig2011,Sun2015}, leading to long-range superconducting proximity~effects~\cite{Duckheim2011,Bergeret2013,Bergeret2014,Jacobsen2015}, and Majorana states~\cite{Nilsson2008,Duckheim2011,Lee2012a,Nadj-Perge2014,Dumitrescu2015,Pawlak2016,Ruby2017,Livanas2019}, which are expected to form in superconducting proximity regions in the presence of SOC.
    
        While SOC in bulk materials plays the key role for \emph{intrinsic anomalous~Hall~effects}~\cite{Hall1881,Wolfle2006,Nagaosa2006,Sinitsyn2008,Nagaosa2010}, recent theoretical~studies~\cite{Vedyayev2013,Vedyayev2013a,MatosAbiague2015,HuongDang2015,HuongDang2018,Zhuravlev2018} predicted that interfacial~SOC in F/normal~metal~(N) tunnel~junctions can give rise to \emph{extrinsic tunneling~anomalous~Hall~effects~(TAHEs)} in the N, owing to spin-polarized \emph{skew~tunneling} of electrons through the interface. The unique scaling of the associated TAHE~conductances could make the effect a fundamental tool for identifying and characterizing interfacial SOC, thus providing the input for tailoring systems that could, e.g., host Majoranas. Although first experiments on granular junctions~\cite{Rylkov2017} confirmed the predictions, the extremely small TAHE~conductances remain one of the main obstacles. Sizable TAHE~conductances require either interfacial barriers with large~SOC, such as ferroelectric~semiconductors~(SCs)~\cite{Zhuravlev2018}, or different junction~compositions.
    
        In this Rapid~Communication, we consider F/SC/S~junctions, in which the N~electrode is replaced by a S.
        We demonstrate that, analogously to the tunneling~picture in the normal-conducting case, \emph{skew~reflection}~\footnote{Conventional \emph{skew scattering} actually refers to momentum- and spin-dependent scattering of spin-polarized charge carriers on magnetic impurities. To clearly differentiate between that and our reflection-based mechanism (which does not require the presence of impurities at all), we rely on the term \emph{skew reflection}.} of spin-polarized carriers at the barrier leads to TAHEs in the F. Due to the presence of a S~electrode, we distinguish two skew reflection processes: \emph{skew specular reflection~(SR) and skew Andreev reflection~(AR)}. By formulating a qualitative physical picture including both processes, we assert that skew~SR and skew~AR can act together and significantly enhance the TAHE compared to all previously studied (normal) systems. Special attention must be paid to skew~AR, which transfers Cooper~pairs across the barrier into the S. The electrons forming one Cooper~pair are thereby also subject to the proposed skew~reflection mechanism. We discuss that the result is a \emph{transverse supercurrent response}, initially deduced from a phenomenological Ginzburg-Landau~treatment~\cite{Mironov2017}, \emph{with widely tunable characteristics}. Both findings, relatively giant TAHE~conductances in the F and transverse supercurrents in the S, are distinct fingerprints to experimentally detect skew~AR and characterize the junctions' interfacial~SOC.

\begin{figure}
        	\includegraphics[width=0.46\textwidth]{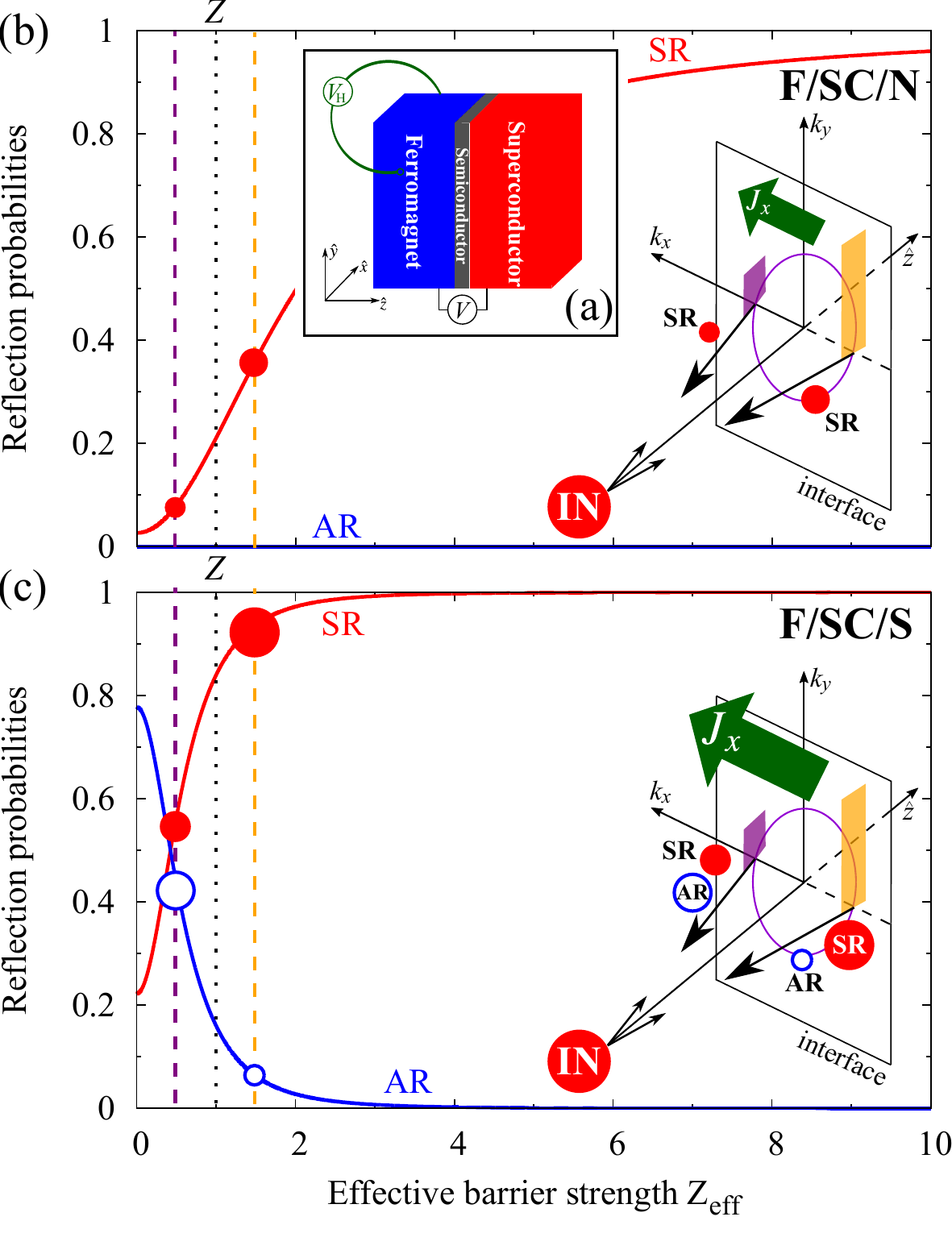}
        	\colorcaption{(a)~Sketch of the considered F/SC/S~junction, using $ C_{2v} $ principal crystallographic orientations, $ \hat{x} \parallel [110] $, $ \hat{y} \parallel [\overline{1}10] $, and $ \hat{z} \parallel [001] $. 
        	(b)~Calculated (zero-bias) \emph{normal-state} reflection probabilities for incident spin~up electrons~(IN) at the SC interface, invoking AR and SR, as a function of~$ Z_\mathrm{eff} = (2mV_\mathrm{eff})/(\hbar^2 k_\mathrm{F}) = Z - (2 \sigma m \alpha k_x)/(\hbar^2 k_\mathrm{F}) $ [dimensionless Blonder-Tinkham-Klapwijk~(BTK)-like barrier parameter for the effective scattering potential in Eq.~\eqref{EqEffPotential}]; $ Z = (2mV_0d)/(\hbar^2 k_\mathrm{F}) $ (see black dashed line for an example) is the usual (spin-independent) barrier strength. Owing to skew~reflection, electrons with $ k_x < 0 $ are exposed to an effectively lowered (dashed violet line) and those with $ k_x >0 $ to a raised (dashed orange line) barrier; the carrier imbalance (carrier densities are proportional to the size of the red and blue circles)  generated via skew SR generates then the transverse Hall current~$ J_x $ (voltage drop~$ V_\mathrm{H} $). The skew reflection mechanism is schematically illustrated in the inset.
        	(c)~Same as in~(b), but for the \emph{superconducting} scenario, in which additionally skew AR plays a key role.
        	    \label{Fig1}}
        \end{figure}

        We consider a biased ballistic F/SC/S~junction grown along the $ \hat{z} $-direction, in which the two semi-infinite F and S regions are separated by an ultrathin SC~barrier~[see Fig.~\ref{Fig1}(a)].
        The barrier may be composed of a thin layer of zincblende materials~(e.g., GaAs or InAs) and introduces potential scattering, as well as strong interfacial Rashba~\cite{Bychkov1984} and Dresselhaus~\cite{Dresselhaus1955} SOC~\cite{Fabian2004,Fabian2007}.
        
        The system can be modeled by means of the stationary Bogoljubov--de~Gennes~(BdG) Hamiltonian~\cite{DeGennes1989},
        \begin{equation}
            \hat{\mathcal{H}}_\mathrm{BdG} = \left[ \begin{matrix} \hat{\mathcal{H}}_\mathrm{e} & \hat{\Delta}_\mathrm{S}(z) \\ \hat{\Delta}_\mathrm{S}^\dagger(z) & \hat{\mathcal{H}}_\mathrm{h} \end{matrix} \right] ,
        \end{equation}
        where $ \hat{\mathcal{H}}_\mathrm{e}=[-\hbar^2/(2m) \, \boldsymbol{\nabla}^2-\mu] \, \hat{\sigma}_0 - (\Delta_\mathrm{XC}/2) \, \Theta(-z) \, (\hat{\mathbf{m}} \cdot \hat{\boldsymbol{\sigma}}) + V_\mathrm{SC} \, d_\mathrm{SC} \, \hat{\sigma}_0 \, \delta(z) + \hat{\mathcal{H}}_\mathrm{SC}^\mathrm{SOC} \, \delta(z) $ represents the single-electron Hamiltonian and $ \hat{\mathcal{H}}_\mathrm{h} = -\hat{\sigma}_y \, \hat{\mathcal{H}}_\mathrm{e}^* \, \hat{\sigma}_y $ its holelike counterpart ($ \hat{\sigma}_0 $ and $ \hat{\sigma}_i $ indicate the $ 2 \times 2 $~identity and the $ i $th Pauli matrix; $ \hat{\boldsymbol{\sigma}} = [\hat{\sigma}_x, \, \hat{\sigma}_y, \, \hat{\sigma}_z]^\top $ is the vector of Pauli matrices).
        The F is described within the Stoner model with exchange energy $ \Delta_\mathrm{XC} $ and magnetization direction $ \hat{\mathbf{m}} = [\cos \Phi, \, \sin \Phi, \, 0]^\top $, where $ \Phi $ is measured with respect to the $ \hat{x} $-axis. Following earlier studies~\cite{DeJong1995,Zutic1999,Zutic2000,Costa2017,Costa2018}, the ultrathin SC~layer is included into our model as a $ \delta $-like barrier with height $ V_\mathrm{SC} $ and width $ d_\mathrm{SC} $; its SOC enters the Hamiltonian~\cite{Fabian2004,Fabian2007} $ \hat{\mathcal{H}}_\mathrm{SC}^\mathrm{SOC} = \alpha \, (k_y \, \hat{\sigma}_x - k_x \, \hat{\sigma}_y) - \beta \, (k_y \, \hat{\sigma}_x + k_x \, \hat{\sigma}_y) $, where the first part accounts for SOC of the Rashba type and the second part resembles linearized Dresselhaus SOC~\footnote{Higher-order SOC~terms can be neglected since the main current~contributions come from states with small~$ k_x $ and~$ k_y $~\cite{MatosAbiague2009}}, both with the effective strengths $ \alpha $ and $ \beta $, respectively.
        Inside the S electrode, the S~pairing~potential, $ \hat{\Delta}_\mathrm{S}(z) = |\Delta_\mathrm{S}| \, \Theta(z) $ ($ |\Delta_\mathrm{S}| $ is the isotropic energy gap of the S), couples the electron and hole blocks of the BdG~Hamiltonian. Note that although writing $ \hat{\Delta}_\mathrm{S} $ in that way is a rigid approximation, neglecting proximity~effects, this approach still yields reliable results for transport calculations~\cite{Likharev1979,Beenakker1997}.
        For the sake of simplicity, we further assume the same Fermi levels, $ \mu $, and equal effective carrier masses, $ m $, in the F and S.
        
        Assuming translational invariance parallel to the barrier, the solutions of the BdG~equation, $ \hat{\mathcal{H}}_\mathrm{BdG} \, \Psi^\sigma(\mathbf{r}) = E \, \Psi^\sigma(\mathbf{r}) $, can be factorized according to $ \Psi^\sigma(\mathbf{r}) = \psi^\sigma(z) \, \mathrm{e}^{\mathrm{i} \, (\mathbf{k_\parallel} \cdot \mathbf{r_\parallel})} , $ where $ \mathbf{k_\parallel} = [k_x, \, k_y, \, 0]^\top $ ($ \mathbf{r_\parallel}=[x, \, y, \, 0]^\top $) denotes the in-plane momentum (position) vector and $ \psi^\sigma(z) $ are the BdG~equation's individual solutions for the reduced one-dimensional scattering problem along $ \hat{z} $. The latter account for the different involved scattering processes at the SC interface: incoming electrons with spin~$ \sigma $ [$ \sigma=+(-)1 $ for spin~up~(down), which effectively indicates a spin parallel~(antiparallel) to $ \hat{\mathbf{m}} $] may either undergo AR or SR, or may be transmitted as quasiparticles into the S.

        Due to the presence of interfacial SOC, electrons incident on the ultrathin SC are exposed to an effective scattering potential that incorporates besides the usual barrier strength (determined by the barrier's height and width) also the in-plane momentum- and spin-dependent contribution of the SOC. To extract valuable qualitative trends from our model, we first focus on the simple situation in which only Rashba~SOC is present~($ \alpha > 0 $, $ \beta = 0 $), the F's magnetization is aligned along~$ \hat{y} $ ($ \Phi = \pi/2 $), and $ k_y=0 $. In this case, the effective scattering potential reads
        \begin{equation}
            V_\mathrm{eff} = V_\mathrm{SC} \, d_\mathrm{SC} - \sigma \, \alpha \, k_x ,
            \label{EqEffPotential}
        \end{equation}
        where the first part represents the usual barrier strength and the second the SOC-dependent part. Assuming that SOC is weak and spin-flip scattering becomes negligible, only spin-conserving AR and SR are allowed inside the F, each with certain probabilities. The latter, extracted from an extended Blonder--Tinkham--Klapwijk~(BTK) model~\cite{Blonder1982} by substituting the effective scattering potential in Eq.~\eqref{EqEffPotential}~[see the Supplemental Material (SM)~\footnote{See the attached Supplemental~Material, including Refs.~\cite{Bychkov1984,Dresselhaus1955,Fabian2004,Fabian2007,DeJong1995,Zutic1999,Zutic2000,Costa2017,Costa2018,DeGennes1989,Blonder1982,Hoegl2015,*Hogl2015,MatosAbiague2015,Rylkov2017,Furusaki1991,McMillan1968,Moser2007,MatosAbiague2009,Nitta1997,Koga2002,Chen2018,Giancoli1995,Wang2006,*Wang2007,Zhuravlev2018}, for more details.} for details], are shown for incoming spin~up electrons as a function of~$ V_\mathrm{eff} $ in Figs.~\ref{Fig1}(b) and (c), once for the normal state and once for the superconducting junction.
        
        In the first case, AR is completely forbidden, while the probability that the incident electron gets specularly reflected continuously increases with increasing effective scattering potential; note that there is also a finite transmission probability into the right normal-state electrode (not shown). For a constant \emph{moderate} barrier height and width (black dashed line) and nonzero Rashba~SOC, Eq.~\eqref{EqEffPotential} suggests that incoming spin~up electrons with positive~$ k_x $ experience a significantly lower barrier (violet dashed line) and thus undergo \emph{skew~SR} with a lower probability than those with negative~$ k_x $~(orange dashed line). The generated spatial charge imbalance in the F must be compensated by a transverse Hall current flow, $ J_x $, along~$ \hat{x} $. Strictly speaking, the situation gets reversed for incident spin~down electrons. Nevertheless, since there are more occupied spin~up states, both channels cannot completely cancel and a finite Hall current remains.
        
        If the junction becomes superconducting, AR comes into play. Although the AR~probability generally decreases with increasing~$ V_\mathrm{eff} $, the crucial point is that AR involves holes. Consequently, \emph{skew~AR} produces simultaneously an electron~excess also at negative~$ k_x $, and both skew~AR and SR act together to noticeably increase the transverse Hall~current.
        
        Another important observation relies on the reflection~probabilities at large~$ V_\mathrm{eff} $. In both junction scenarios, the SR probabilities approach unity at $ V_\mathrm{eff} \gg (\hbar^2 k_\mathrm{F}) / (2m) $; this happens much faster in the superconducting case than in the normal state. The scattering potential is then mostly determined by the usual barrier height and width, and the spin-dependent contribution only barely impacts~$ V_\mathrm{eff} $. Therefore, both skew reflection and the Hall~currents are expected to be strongly damped in the presence of \emph{strong} barriers, in superconducting even more than in normal-conducting junctions.
        
        As a clear fingerprint to experimentally detect skew AR, our qualitative picture suggests a significant enhancement of the superconducting junctions' TAHE~conductance, compared to the normal-state regime. To evaluate the TAHE~conductances along the transverse~$ \hat{\eta} $-direction ($ \eta \in \{ x;y \} $), we follow a generalized BTK~approach~\cite{Blonder1982}, yielding the \emph{zero-temperature} TAHE~conductances
            \begin{multline}
                G_{\eta,z} = -\frac{G_0 A}{8\pi^2} \sum_{\sigma = \pm 1} \int \mathrm{d}^2 \mathbf{k_\parallel} \, \frac{k_\eta}{k_z^\sigma} \, \left\{ \left[ \left| r^{\sigma,\sigma}_\mathrm{e}(eV) \right|^2 + \left| r^{\sigma,-\sigma}_\mathrm{e}(eV) \right|^2 \right] \right. \\
                \left. + \left[ \left| r^{\sigma,-\sigma}_\mathrm{h}(-eV) \right|^2 + \left| r^{\sigma,\sigma}_\mathrm{h}(-eV) \right|^2 \right] \right\} ,
                \label{EqTransverseConductance}
            \end{multline}
        where $ G_0 = (2e^2) / h $ abbreviates the conductance quantum, $ A $ stands for the cross-section area, $ k_z^\sigma = \sqrt{k_\mathrm{F}^2 \, (1+\sigma P) - \mathbf{k}^2_\mathbf{\parallel}} $ represents the $ \hat{z} $-component of the particles' wave vector in the~F with spin~polarization~$ P = (\Delta_\mathrm{XC}/2)/\mu $, and $ k_\mathrm{F} = \sqrt{2m\mu}/\hbar $ is the Fermi wave vector. The reflection coefficients $ r_\mathrm{e}^{\sigma,\sigma} $ ($ r_\mathrm{e}^{\sigma,-\sigma} $) correspond to SR (spin-flip SR), while $ r_\mathrm{h}^{\sigma,-\sigma} $ ($ r_\mathrm{h}^{\sigma,\sigma} $) indicate AR (spin-flip AR). Unlike for the (longitudinal) tunneling conductance~\cite{Blonder1982}, SR and AR contribute to the Hall conductances with the \emph{same} sign since the specularly reflected electron and the Andreev reflected hole move into opposite transverse directions; the different sign in the transverse velocities gets then compensated by the opposite charge of electrons and holes. Therefore, the charge imbalances created by skew~AR and SR can indeed give rise to individual Hall~currents that flow along the same direction, and finally lead to sizable Hall~responses in superconducting junctions.
        \begin{figure}
        	\includegraphics[width=0.475\textwidth]{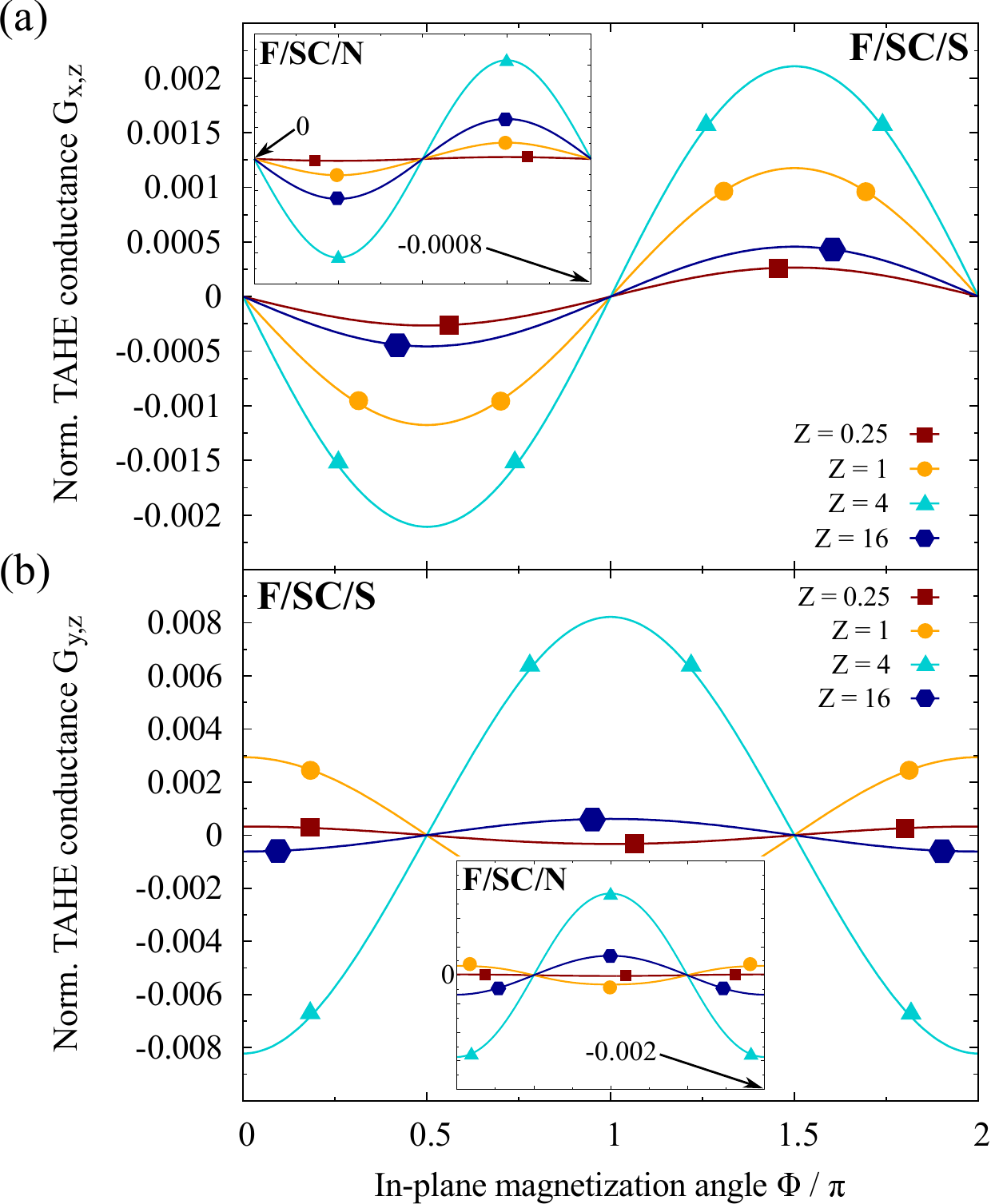}
            \colorcaption{Calculated dependence of the zero-bias TAHE~conductances, (a)~$ G_{x,z} $ and (b)~$ G_{y,z} $, normalized to Sharvin's conductance, $ G_\mathrm{S}=(Ae^2 k_\mathrm{F}^2)/(4\pi^2 h) $, on the in-plane magnetization angle, $ \Phi $, and for various indicated barrier strengths, $ Z $; the SOC parameters are~$ \alpha = 42.3 \, \mathrm{eV} \, \text{\AA}^2 $ and~$ \beta \approx 19.2 \, \mathrm{eV} \, \text{\AA}^2 Z $. The insets show similar normal state calculations, when the S is replaced by a N.
                \label{Fig2}}
         \end{figure}
       
        To elaborate on the TAHE~conductances' main features, we evaluate Eq.~\eqref{EqTransverseConductance} for Fe/GaAs/V~like model junctions. The spin~polarization in~Fe is $ P=0.7 $ (Fermi wave vector $ k_\mathrm{F} \approx 8 \times 10^7 \, \mathrm{cm}^{-1} $~\cite{Martinez2018}), while $ |\Delta_\mathrm{S}| \sim 1.6 \, \mathrm{meV} $ refers to V's gap~\cite{Martinez2018}. The (material-specific) Dresselhaus SOC~strength of GaAs can be approximated~\cite{Fabian2007,MatosAbiague2009} as $ \beta \approx Z k_\mathrm{F} \gamma $, with $ \gamma \approx 24 \, \mathrm{eV} \, \text{\AA}^3 $ being the cubic Dresselhaus parameter for GaAs~\cite{Fabian2007}. The GaAs barrier's height and width are captured by the dimensionless BTK-like barrier measure~$ Z = (2m V_\mathrm{SC} d_\mathrm{SC}) / (\hbar^2 k_\mathrm{F}) $ (typically, $ V_\mathrm{SC} \sim 0.75 \, \mathrm{eV} $~\cite{MatosAbiague2009} so that $ Z = 1 $ represents a barrier with thickness $ d_\mathrm{SC} \sim 0.40 \, \mathrm{nm} $). Figure~\ref{Fig2} shows the dependence of the normalized zero-bias~\footnote{Experimental measurements of the TAHE~response in the F simultaneously detect a contribution stemming from conventional anomalous~Hall~effects. To separate both parts, one could exploit the TAHE~contribution's unique voltage dependence~\cite{Note3}.} TAHE~conductances, $ G_{x,z} $ and $ G_{y,z} $, on the F's~magnetization~orientation for various barrier strengths, $ Z $, and the Rashba~SOC parameter~$ \alpha \approx 42.3 \, \mathrm{eV} \, \text{\AA}^2 $, which lies well within the experimentally accessible values~\cite{Moser2007,MatosAbiague2009}. To quantitatively compare the conductance amplitudes, the insets show analogous calculations in the normal-conducting state.
        
        \begin{figure}
            \includegraphics[width=0.475\textwidth]{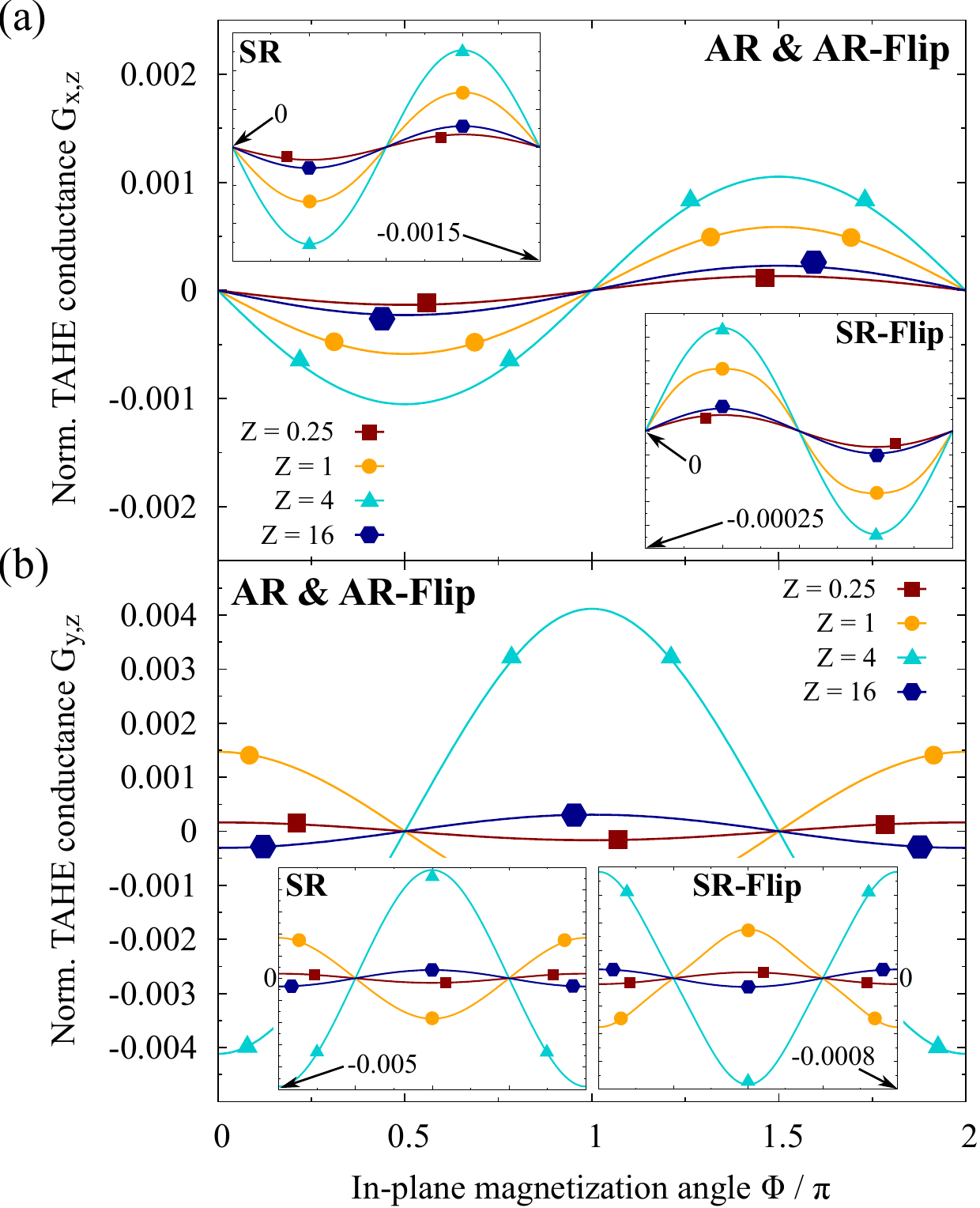} 
            \colorcaption{Calculated dependence of the zero-bias TAHE~conductances, (a)~$ G_{x,z} $ and (b)~$ G_{y,z} $, normalized to~$ G_\mathrm{S} $, on the in-plane magnetization angle, $ \Phi $, for the same parameters as in Fig.~\ref{Fig2}. The contributions stemming from SR and spin-flip~SR (SR-Flip), and similarly those originating from ARs, are separately resolved.
              \label{Fig3}}
        \end{figure}
        
        Our simulations reveal all the TAHE~conductances' important properties. First, we observe the sinelike (cosinelike) variation of $ G_{x,z} $ ($ G_{y,z} $) with respect to the F's magnetization angle. Those dependencies follow from symmetry~considerations~\cite{Note3} and unambiguously reflect the junction's magnetoanisotropic transport characteristics~\cite{MatosAbiague2015}. 
        Second, we find  that skew AR and SR can indeed act together in superconducting junctions, leading to sizable TAHE~conductances (and voltages~\cite{Note3}). Specifically, $ G_{x,z} $ can be increased by more than one order of magnitude and $ G_{y,z} $ still roughly by a factor of~4, compared to normal~junctions. 
        However, the full physical mechanism is more complicated than our simple picture in~Fig.~\ref{Fig1}, where we considered one particular combination of in-plane momenta. To obtain the total TAHE~conductances, we need to average over all possible configurations~[see Eq.~\eqref{EqTransverseConductance}], which can---mostly depending on the barrier and Rashba~SOC strengths---also reverse the Hall~current's direction, observed, for~example, in~$ G_{y,z} $ by increasing $ Z $ from $ Z=1 $ to $ Z=4 $~\cite{Note3}.
        Finally, we can confirm the stated connection between the skew reflection mechanism and the TAHE~conductances for strong tunneling barriers. As~$ Z $ increases, $ V_\mathrm{eff} $ is mostly determined by the bare barrier strength itself~[see Eq.~\eqref{EqEffPotential}], and the momentum- and spin-dependent SOC asymmetry, responsible for the Hall current generation, gets remarkably suppressed (especially in the superconducting regime). As a result, strong barriers significantly decrease the TAHE~conductances.

        To resolve AR and SR, Fig.~\ref{Fig3} shows their spin-resolved conductance contributions. The spin-flip AR part is not separately shown as its amplitudes are up to two orders of magnitude smaller than those of (spin-conserving) AR. Interestingly, the total TAHE~conductance is nearly fully dominated by (spin-conserving) AR and SR; both contributions are comparable in magnitude and have the same signs so that they indeed add up, resulting in sizable TAHE~conductances. Since spin-flip SR involves electrons with opposite spin, the effective barrier picture in Fig.~\ref{Fig1} gets reversed and the related TAHE~conductance contribution changes sign. Nevertheless, this contribution is much smaller than those attributed to spin-conserving skew reflections so that it cannot modify the TAHE~conductances' qualitative features.

        \begin{figure}
        	\includegraphics[width=0.475\textwidth]{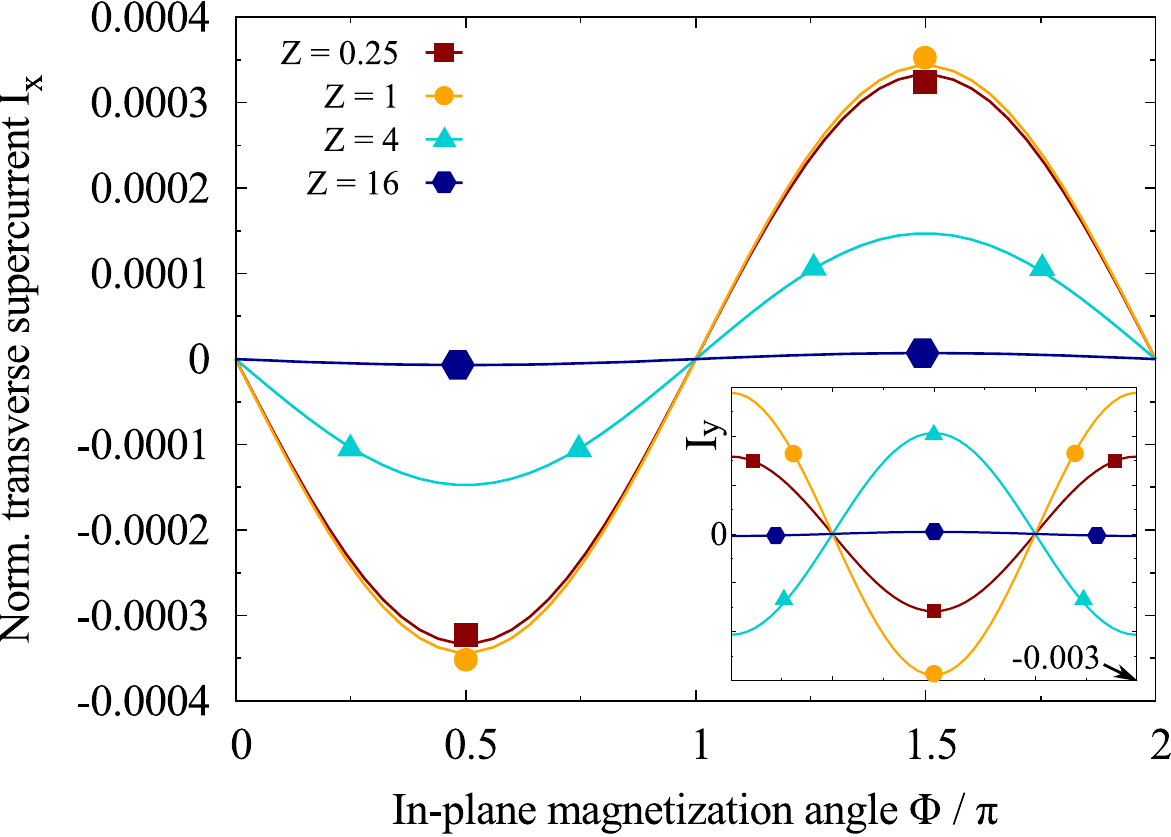}
            \colorcaption{Calculated dependence of the zero-bias transverse supercurrent response, $ I_x $ ($ I_y $ in the inset), normalized according to $ [I_{x \, (y)} e]/(G_\mathrm{S}\pi|\Delta_\mathrm{S}|) $, on the in-plane magnetization angle, $ \Phi $, for the same parameters as in Fig.~\ref{Fig2}.
               \label{Fig4}}
        \end{figure}
        AR is \emph{the} crucial scattering process at metal/S~interfaces; it transfers Cooper~pairs, converting normal into supercurrents, plays an important role in experimentally quantifying Fs' spin~polarization~\cite{Soulen1999}, and is also essential for the sizable TAHE~conductances in the F of our  system. Particularly interesting are the transferred Cooper~pairs, which are also exposed to the effective scattering potential and may thus trigger a response in terms of a transverse supercurrent in the S~\cite{Mironov2017}. Within our model, we evaluate the zero-bias supercurrent~components, $ I_\eta $, starting from a generalized Furusaki--Tsukada~technique~\cite{Furusaki1991}~(see the SM~\cite{Note3}). For the considered parameters, we stated that the main skew AR~contribution to the TAHE~conductance comes from the spin-conserving process. As the latter involves spin-singlet Cooper~pairs, composed of electrons with opposite transverse momenta and spins, one could generalize our skew reflection picture to a combined one for the two individual Cooper~pair electrons. As a consequence, the induced supercurrents' qualitative features follow the same trends as those of the TAHE~conductances in Fig.~\ref{Fig2}. 
        Figure~\ref{Fig4}, presenting $ I_\eta $ as a function of the magnetization angle, $ \Phi $, confirms this expectation: 
        the supercurrent components' dependence on~$ \Phi $ and their orientations (signs) reflect one-to-one the properties of the TAHE~conductances in the F. Even the sign change we explored in~$ G_{y,z} $ when changing~$ Z $ from~$ Z=1 $ to~$ Z=4 $ is (qualitatively) transferred into the supercurrent response~$ I_y $. Nevertheless, there is one important difference from the TAHE~conductance, concerning the currents' magnitudes. The supercurrent~response always results from two single electrons that tunnel into the~S, forming a Cooper~pair. In order to generate sizable supercurrents, both electrons must simultaneously skew tunnel into the S (mediated by skew~AR), which is less likely to happen at strong barriers than skew~tunneling of unpaired electrons. Therefore, the maximal supercurrent amplitudes---several milliamperes for optimal configurations---occur at smaller $ Z $ than the maximal TAHE~conductance amplitudes in the F.

        To conclude, we investigated the intriguing consequences of skew~AR and SR at SC~interfaces of superconducting tunnel junctions. We predict that the interplay of both skew~reflection processes can constructively amplify their effects. Furthermore, also the Cooper~pairs transferred into the S via AR~cycles are subject to interfacial skew reflections. As a result, both sizable TAHE conductances in the F and characteristically modulating transverse supercurrents in the S are generated, opening new venues for experimental and theoretical studies.
    
    \begin{acknowledgments}
        This work was supported by the International~Doctorate~Program Topological~Insulators of the Elite~Network of Bavaria and DFG~SFB Grant~No.~1277, project B07 (A.C. and J.F.), as well as by DARPA Grant~No.~DP18AP900007 and US~ONR Grant~No.~N000141712793 (A.M.-A.).
    \end{acknowledgments}

    \bibliography{paper}

    \onecolumngrid
    \newpage
    

\section*{Supplementary material}

# SUPPLEMENTAL MATERIAL
# Skew Andreev reflection in ferromagnet/superconductor junctions

Andreas Costa,[1,*] Alex Matos-Abiague,[2] and Jaroslav Fabian[1]

[1]*Institute for Theoretical Physics, University of Regensburg, 93040 Regensburg, Germany*
[2]*Department of Physics and Astronomy, Wayne State University Detroit, Michigan 48201, USA*

In this Supplemental Material, we present the computational details not included into the manuscript and clarify the impact of the remaining parameters, which could be particularly important in experiments to identify suitable parameter regimes for a reliable measurement. We use the abbreviations declared in our manuscript and the same junction parameters if not specifically indicated.

## I. BDG SCATTERING STATES IN THE F AND S REGIONS

As introduced in the manuscript, we consider a three-dimensional tunnel junction composed of a semi-infinite F region and a semi-infinite S region, spanning $z < 0$ and $z > 0$ half-spaces, respectively. Both electrodes are separated by an ultrathin SC, which breaks the space inversion symmetry and simultaneously gives rise to interfacial Rashba [S1] and Dresselhaus [S2] SOC [S3, S4]. We include that layer into our model in terms of a deltalike potential barrier [S5–S9]. In particle-hole Nambu space, relying on the basis set $\Psi = \left[\psi^\uparrow, \psi^\downarrow, (\psi^\downarrow)^\dagger, (-\psi^\uparrow)^\dagger\right]^\top$, the junction's stationary BdG equation [S10] reads

$$\begin{bmatrix} \hat{\mathcal{H}}_e & \hat{\Delta}_S(z) \\ \hat{\Delta}_S^\dagger(z) & \hat{\mathcal{H}}_h \end{bmatrix} \Psi^\sigma(\mathbf{r}) = E\, \Psi^\sigma(\mathbf{r}). \tag{S1}$$

The single-particle Hamiltonians for electrons and holes, $\hat{\mathcal{H}}_e$ and $\hat{\mathcal{H}}_h$, as well as the S pairing potential, $\hat{\Delta}_S(z)$, are defined in the manuscript.

As particles are solely scattered along the $\hat{z}$-direction, we can assume the wave vector parallel to the interface, $\mathbf{k}_\parallel = \left[k_x, k_y, 0\right]^\top$, to be conserved. Therefore, in order to obtain the BdG equation's eigenstates, we substitute the general ansatz

$$\Psi^\sigma(\mathbf{r}) = \psi^\sigma(z)\, e^{i(\mathbf{k}_\parallel \cdot \mathbf{r}_\parallel)}, \tag{S2}$$

with $\mathbf{r}_\parallel = [x, y, 0]^\top$, into Eq. (S1), to effectively reduce our problem to finding the one-dimensional scattering states $\psi^\sigma(z)$ for a given spin $\sigma$ [$\sigma = +(-)1$ for spin parallel (antiparallel) to the magnetization unit vector, $\hat{\mathbf{m}}$, in the F; note that we simultaneously also use the established terms spin up and spin down for $\sigma = 1$ and $\sigma = -1$].

For an incoming electron with spin $\sigma$ from the F electrode, $\psi^\sigma(z)$ is found to be

$$\psi^\sigma(z<0) = \frac{1}{\sqrt{2}} \begin{bmatrix} 1 \\ \sigma e^{i\Phi} \\ 0 \\ 0 \end{bmatrix} e^{ik_{z,e}^\sigma z} + r_e^{\sigma,\sigma} \frac{1}{\sqrt{2}} \begin{bmatrix} 1 \\ \sigma e^{i\Phi} \\ 0 \\ 0 \end{bmatrix} e^{-ik_{z,e}^\sigma z} + r_e^{\sigma,-\sigma} \frac{1}{\sqrt{2}} \begin{bmatrix} 1 \\ -\sigma e^{i\Phi} \\ 0 \\ 0 \end{bmatrix} e^{-ik_{z,e}^{-\sigma} z}$$

$$+ r_h^{\sigma,-\sigma} \frac{1}{\sqrt{2}} \begin{bmatrix} 0 \\ 0 \\ 1 \\ \sigma e^{i\Phi} \end{bmatrix} e^{ik_{z,h}^{-\sigma} z} + r_h^{\sigma,\sigma} \frac{1}{\sqrt{2}} \begin{bmatrix} 0 \\ 0 \\ 1 \\ -\sigma e^{i\Phi} \end{bmatrix} e^{ik_{z,h}^\sigma z}. \tag{S3}$$

As we mentioned in the manuscript, the SR coefficients without and with a spin-flip are denoted by $r_e^{\sigma,\sigma}$ and $r_e^{\sigma,-\sigma}$ (the first $\sigma$ refers to the incoming electron's spin and the second one to the spin of the reflected electron), while we use the coefficients $r_h^{\sigma,-\sigma}$ and $r_h^{\sigma,\sigma}$ for AR and spin-flip AR. Note that in the ordinary AR process (without a spin-flip), the retroreflected hole has by definition opposite spin compared to the incident electron. Analogously, $\psi^\sigma(z)$ in the S electrode can be written as

$$\psi^\sigma(z>0) = t_e^{\sigma,\sigma} \begin{bmatrix} u \\ 0 \\ v \\ 0 \end{bmatrix} e^{iq_{z,e}z} + t_e^{\sigma,-\sigma} \begin{bmatrix} 0 \\ u \\ 0 \\ v \end{bmatrix} e^{iq_{z,e}z} + t_h^{\sigma,\sigma} \begin{bmatrix} v \\ 0 \\ u \\ 0 \end{bmatrix} e^{-iq_{z,h}z} + t_h^{\sigma,-\sigma} \begin{bmatrix} 0 \\ v \\ 0 \\ u \end{bmatrix} e^{-iq_{z,h}z}, \tag{S4}$$

---

* Corresponding author: andreas.costa@physik.uni-regensburg.de

accounting for transmission of the incident electron as an electronlike quasiparticle (coefficients $t_\text{e}^{\sigma,\pm\sigma}$) or as a holelike quasiparticle (coefficients $t_\text{h}^{\sigma,\pm\sigma}$). The $\hat{z}$-components of the electrons' and holes' wave vectors in the F are given by

$$k_{z,\text{e}}^\sigma = \sqrt{\frac{2m}{\hbar^2}\left[\mu + E + \sigma\frac{\Delta_\text{XC}}{2}\right] - \mathbf{k}_\|^2} \quad \text{and} \quad k_{z,\text{h}}^\sigma = \sqrt{\frac{2m}{\hbar^2}\left[\mu - E + \sigma\frac{\Delta_\text{XC}}{2}\right] - \mathbf{k}_\|^2}, \tag{S5}$$

while those in the S are

$$q_{z,\text{e}} = \sqrt{\frac{2m}{\hbar^2}\left[\mu + \sqrt{E^2 - |\Delta_\text{S}|^2}\right] - \mathbf{k}_\|^2} \quad \text{and} \quad q_{z,\text{h}} = \sqrt{\frac{2m}{\hbar^2}\left[\mu - \sqrt{E^2 - |\Delta_\text{S}|^2}\right] - \mathbf{k}_\|^2}. \tag{S6}$$

In a typical situation, the (quasi)particle excitation energies, $E$, as well as the superconducting gap, $|\Delta_\text{S}|$, are both much smaller than the chemical potential, i.e., $E \ll \mu$ and $|\Delta_\text{S}| \ll \mu$. Therefore, we can use the commonly approximated wave vectors $k_{z,\text{e}}^\sigma \approx k_{z,\text{h}}^\sigma \approx k_z^\sigma = \sqrt{k_\text{F}^2(1 + \sigma P) - \mathbf{k}_\|^2}$ [$P = (\Delta_\text{XC}/2)/\mu$ is an effective measure for the F's spin polarization] and $q_{z,\text{e}} \approx q_{z,\text{h}} \approx q_z = \sqrt{k_\text{F}^2 - \mathbf{k}_\|^2}$ to dramatically simplify the further theoretical treatment of our junction; $k_\text{F} = \sqrt{2m\mu}/\hbar$ is the Fermi wave vector, which we assumed to be the same in all constituents. The factors $u = u(E)$ and $v = v(E)$, appearing in the S's scattering states, are the usual BCS coherence factors, satisfying

$$u(E) = \sqrt{\frac{1}{2}\left(1 + \sqrt{1 - \frac{|\Delta_\text{S}|^2}{E^2}}\right)} = \sqrt{1 - v^2(E)}. \tag{S7}$$

To obtain the so far unknown scattering coefficients—which generally depend on the particle energy via the coherence factors and on the parallel momentum via the wave vectors—the states need to fulfill the interfacial ($z = 0$) boundary conditions

$$\psi^\sigma(z)\big|_{z=0_-} = \psi^\sigma(z)\big|_{z=0_+}, \tag{S8}$$

$$\left\{\left[\frac{\hbar^2}{2m}\frac{\text{d}}{\text{d}z} + V_\text{SC}d_\text{SC}\right]\boldsymbol{\eta}\psi^\sigma(z)\right\}\bigg|_{z=0_-} + \begin{bmatrix}\boldsymbol{\Omega}\cdot\hat{\boldsymbol{\sigma}} & \mathbf{0}_{2\times 2} \\ \mathbf{0}_{2\times 2} & -(\boldsymbol{\Omega}\cdot\hat{\boldsymbol{\sigma}})^*\end{bmatrix}\psi^\sigma(z)\bigg|_{z=0_-} = \frac{\hbar^2}{2m}\frac{\text{d}}{\text{d}z}\boldsymbol{\eta}\psi^\sigma(z)\bigg|_{z=0_+}, \tag{S9}$$

where $\boldsymbol{\eta} = \text{diag}[\hat{\sigma}_0, -\hat{\sigma}_0]$; $\boldsymbol{\Omega} = [\alpha k_y - \beta k_x, -\alpha k_x - \beta k_x, 0]$ contains the Rashba and Dresselhaus spin-orbit fields.

The resulting eight-dimensional system of equations can then be numerically solved for a generic bias voltage $V$ by setting the excitation energy $E = eV$. However, at zero bias ($E = eV = 0$), the coherence factors simply reduce to $u(E=0) = 1/\sqrt{2}$ and $v(E=0) = -\text{i}/\sqrt{2}$, and we can easily find analytical expressions for the reflection coefficients. In the (superconducting) F/SC/S junction, those "zero-bias reflection coefficients" read

$$r_\text{e}^{\sigma,\sigma} = \left\{-\overline{\Omega}^2\text{e}^{4\text{i}\Phi}\left(\tilde{k}_z^{-\sigma} - \tilde{k}_z^\sigma\right)\left(\tilde{k}_z^{-\sigma} + \tilde{k}_z^\sigma\right) + 4\text{i}\sigma\tilde{k}_z^{-\sigma}\overline{\Omega}\text{e}^{3\text{i}\Phi}\left[\tilde{q}_z^2 + \left(\tilde{k}_z^\sigma - \text{i}Z\right)^2 + \left|\overline{\Omega}\right|^2\right] + 4\text{i}\sigma\tilde{k}_z^{-\sigma}\overline{\Omega}^*\text{e}^{\text{i}\Phi}\left[\tilde{q}_z^2 + \left(\tilde{k}_z^\sigma - \text{i}Z\right)^2 + \left|\overline{\Omega}\right|^2\right]\right.$$
$$+ \left(\overline{\Omega}^*\right)^2\left[\left(\tilde{k}_z^{-\sigma}\right)^2 - \left(\tilde{k}_z^{-\sigma}\right)^2\right] - 2\text{e}^{2\text{i}\Phi}\left\{2\left[\left(\tilde{k}_z^{-\sigma}\right)^2\left(\tilde{k}_z^\sigma - \text{i}Z\right)^2 - \left(\tilde{q}_z^2 + \text{i}\tilde{k}_z^\sigma Z + Z^2\right)^2\right] - 2\left|\overline{\Omega}\right|^4\right.$$
$$\left.\left.+ \overline{\Omega}\,\overline{\Omega}^*\left[\left(\tilde{k}_z^{-\sigma}\right)^2 - 4\tilde{q}_z^2 - \left(\tilde{k}_z^\sigma - 2\text{i}Z\right)^2\right]\right\}\right\}\bigg/\gamma, \tag{S10}$$

$$r_\text{e}^{\sigma,-\sigma} = \left\{2\tilde{k}_z^\sigma\overline{\Omega}^2\left\{\text{e}^{4\text{i}\Phi}\left(\tilde{k}_z^\sigma - \tilde{k}_z^{-\sigma}\right) + 2\text{i}\sigma\overline{\Omega}\text{e}^{3\text{i}\Phi}\left[\tilde{q}_z^2 + \left(\tilde{k}_z^{-\sigma} - \text{i}Z\right)\left(\tilde{k}_z^\sigma - \text{i}Z\right) + \left|\overline{\Omega}\right|^2\right]\right.\right.$$
$$\left.\left.- 2\text{i}\sigma\overline{\Omega}^*\text{e}^{\text{i}\Phi}\left[\tilde{q}_z^2 + \left(\tilde{k}_z^{-\sigma} - \text{i}Z\right)\left(\tilde{k}_z^\sigma - \text{i}Z\right) + \left|\overline{\Omega}\right|^2\right] + \left(\overline{\Omega}^*\right)^2\left(\tilde{k}_z^{-\sigma} - \tilde{k}_z^\sigma\right)\right\}\right\}\bigg/\gamma, \tag{S11}$$

$$r_\text{h}^{\sigma,-\sigma} = \left\{4\sigma\tilde{k}_z^\sigma\tilde{q}_z\text{e}^{\text{i}\Phi}\left\{\overline{\Omega}\text{e}^{2\text{i}\Phi}\left(\tilde{k}_z^\sigma - \tilde{k}_z^{-\sigma} - 2\text{i}Z\right) + 2\text{i}\sigma\text{e}^{\text{i}\Phi}\left[\tilde{q}_z^2 + \left(\tilde{k}_z^\sigma - \text{i}Z\right)\left(\tilde{k}_z^{-\sigma} + \text{i}Z\right) + \left|\overline{\Omega}\right|^2\right] + \overline{\Omega}^*\left(\tilde{k}_z^\sigma - \tilde{k}_z^{-\sigma} - 2\text{i}Z\right)\right\}\right\}\bigg/\gamma, \tag{S12}$$

$$r_\text{h}^{\sigma,\sigma} = \left\{-8\text{i}\sigma\tilde{k}_z^\sigma\tilde{q}_zZ\text{e}^{\text{i}\Phi}\left(\overline{\Omega}\text{e}^{2\text{i}\Phi} - \overline{\Omega}^*\right)\right\}\bigg/\gamma, \tag{S13}$$

with the common denominator

$$\gamma = \overline{\Omega}^2\text{e}^{4\text{i}\Phi}\left(\tilde{k}_z^{-\sigma} - \tilde{k}_z^\sigma\right)^2 - 4\text{i}\sigma\overline{\Omega}\text{e}^{3\text{i}\Phi}\left(\tilde{k}_z^{-\sigma} - \tilde{k}_z^\sigma\right)\left(\tilde{k}_z^{-\sigma}\tilde{k}_z^\sigma + \tilde{q}_z^2 - Z^2 + \left|\overline{\Omega}\right|^2\right) - 4\text{i}\sigma\overline{\Omega}^*\text{e}^{\text{i}\Phi}\left(\tilde{k}_z^{-\sigma} - \tilde{k}_z^\sigma\right)\left(\tilde{k}_z^{-\sigma}\tilde{k}_z^\sigma + \tilde{q}_z^2 - Z^2 + \left|\overline{\Omega}\right|^2\right)$$
$$+ \left(\overline{\Omega}^*\right)^2\left(\tilde{k}_z^{-\sigma} - \tilde{k}_z^\sigma\right)^2 - 2\text{e}^{2\text{i}\Phi}\left\{2\left[\left(\tilde{k}_z^{-\sigma}\tilde{k}_z^\sigma + \tilde{q}_z^2\right)^2 + \left[\left(\tilde{k}_z^{-\sigma}\right)^2 + \left(\tilde{k}_z^\sigma\right)^2 + 2\tilde{q}_z^2\right]Z^2 + Z^4\right] + 2\left|\overline{\Omega}\right|^4\right.$$
$$\left.- \overline{\Omega}\,\overline{\Omega}^*\left[\left(\tilde{k}_z^{-\sigma}\right)^2 - 6\tilde{k}_z^{-\sigma}\tilde{k}_z^\sigma + \left(\tilde{k}_z^\sigma\right)^2 - 4\tilde{q}_z^2 + 4Z^2\right]\right\}. \tag{S14}$$

Thereby, the dimensionless "tilde wave vectors" are obtained by dividing the above given ones by the Fermi wave vector $k_F$, i.e., $\tilde{k}_z^{\pm\sigma} = k_z^{\pm\sigma}/k_F$ and $\tilde{q}_z = q_z/k_F$; we further omitted the energy argument in the notation of the coefficients since we are only considering the particular case $E = eV = 0$ here. The impact of the tunneling barrier is hidden in the dimensionless BTK-like [S11] parameter $Z = (2mV_{SC}d_{SC})/(\hbar^2 k_F)$, while the effective dimensionless SOC strengths $\bar{\lambda}_R = (2m\alpha)/\hbar^2$ and $\bar{\lambda}_D = (2m\beta)/\hbar^2$ enter the dimensionless SOC "matrix element" $\bar{\Omega} = (\bar{\lambda}_R - \bar{\lambda}_D)k_y/k_F + i(\bar{\lambda}_R + \bar{\lambda}_D)k_x/k_F$. It is easy to check that in the case of absent Rashba and Dresselhaus SOC, $\bar{\Omega} = 0$, the spin-flip reflection coefficients immediately vanish.

In a similar manner, we can analytically compute the reflection coefficients for the (normal-conducting) F/SC/N junction. Within our model, the scattering states in the superconductor reduce to the well-known ones for a N when substituting $u = 1$ and $v = 0$. Solving the system of equations following from the boundary conditions in that particular scenario yields

$$r_e^{\sigma,\sigma} = \left\{\bar{\Omega}e^{2i\Phi}\left(\tilde{k}_z^\sigma + \tilde{k}_z^{-\sigma}\right) + \bar{\Omega}^*\left(\tilde{k}_z^\sigma + \tilde{k}_z^{-\sigma}\right) + 2i\sigma e^{i\Phi}\left[\left(\tilde{k}_z^\sigma - \tilde{q}_z - iZ\right)\left(\tilde{k}_z^{-\sigma} + \tilde{q}_z + iZ\right) - \bar{\Omega}\,\bar{\Omega}^*\right]\right\}/\Gamma, \tag{S15}$$

$$r_e^{\sigma,-\sigma} = \left\{2\tilde{k}_z^\sigma\left(\bar{\Omega}e^{2i\Phi} - \bar{\Omega}^*\right)\right\}/\Gamma, \tag{S16}$$

$$r_h^{\sigma,-\sigma} = 0, \tag{S17}$$

$$r_h^{\sigma,\sigma} = 0, \tag{S18}$$

with the common denominator

$$\Gamma = -\bar{\Omega}e^{2i\Phi}\left(\tilde{k}_z^{-\sigma} - \tilde{k}_z^\sigma\right) + 2i\sigma e^{i\Phi}\left[\left(\tilde{k}_z^{-\sigma} + \tilde{q}_z + iZ\right)\left(\tilde{k}_z^\sigma + \tilde{q}_z + iZ\right) + \left|\bar{\Omega}\right|^2\right] + \bar{\Omega}^*\left(\tilde{k}_z^\sigma - \tilde{k}_z^{-\sigma}\right). \tag{S19}$$

The fact that the AR coefficients become zero in the normal-conducting system is physically clear since AR naturally appears just at metal/S interfaces. As above, the spin-flip SR coefficient, $r_e^{\sigma,-\sigma}$, vanishes in the absence of SOC ($\bar{\Omega} = 0$).

## II. BTK-LIKE PROBABILITIES FOR SKEW SR AND SKEW AR

In the manuscript, we argued that assuming weak SOC justifies to restrict our considerations in the F on spin-conserving AR and SR as the only allowed scattering processes; transmissions are only possible if the S is in the normal-conducting state or at energies above the S gap. Due to the interfacial SOC, incident electrons feel at the SC interface not only the usual potential barrier's strength (determined by the barrier's height and width), but also an additional in-plane momentum- and spin-dependent scattering potential; see Eq. (2) in the manuscript for the limiting case of $\beta = 0$, $k_y = 0$, and $\Phi = \pi/2$. Transferring the resulting effective scattering potential into the dimensionless BTK-like Z-parameter, see Sec. I, i.e., $Z \mapsto Z_{eff} = Z - (2\sigma m\alpha k_x)/(\hbar^2 k_F)$, allows us to write the BTK-like reflection probabilities for SR and AR as

$$R_e^{\sigma,\sigma} = \left|r_e^{\sigma,\sigma}\right|^2 \tag{S20}$$

and

$$R_h^{\sigma,-\sigma} = \sqrt{\frac{1-\sigma P}{1+\sigma P}}\left|r_h^{\sigma,-\sigma}\right|^2, \tag{S21}$$

with

$$r_e^{\sigma,\sigma} = \frac{2\tilde{k}_z^\sigma\left[\tilde{k}_z^{-\sigma}\left(u^2-v^2\right) + \tilde{q}_z\left(u^2+v^2\right) - iZ_{eff}\left(u^2-v^2\right)\right]}{\tilde{k}_z^\sigma\tilde{q}_z\left(u^2+v^2\right) - i\tilde{k}_z^\sigma Z_{eff}\left(u^2-v^2\right) + \tilde{k}_z^{-\sigma}\left[\tilde{k}_z^\sigma\left(u^2-v^2\right) + \tilde{q}_z\left(u^2+v^2\right) + iZ_{eff}\left(u^2-v^2\right)\right] + \tilde{q}_z^2\left(u^2-v^2\right) + Z_{eff}^2\left(u^2-v^2\right)} - 1 \tag{S22}$$

and

$$r_h^{\sigma,-\sigma} = \frac{4\tilde{k}_z^\sigma\tilde{q}_z uv}{\tilde{k}_z^\sigma\tilde{q}_z\left(u^2+v^2\right) - i\tilde{k}_z^\sigma Z_{eff}\left(u^2-v^2\right) + \tilde{k}_z^{-\sigma}\left[\tilde{k}_z^\sigma\left(u^2-v^2\right) + \tilde{q}_z\left(u^2+v^2\right) + iZ_{eff}\left(u^2-v^2\right)\right] + \tilde{q}_z^2\left(u^2-v^2\right) + Z_{eff}^2\left(u^2-v^2\right)}. \tag{S23}$$

The dimensionless "tilde wave vectors" are the same as defined in Sec. I. Figures 1(b)–(c) in the manuscript show the evaluated reflection probabilities for incident spin up electrons ($\sigma = 1$) and zero external bias voltage ($eV = 0$), once for the normal state and once for the superconducting regime. To exemplarily illustrate the effective in-plane momentum- and spin-dependent barrier strengths, responsible for the skew reflection mechanism, we considered $Z = 1$ (black dotted line), $(2m\alpha)/(\hbar^2 k_F) = 1$, and $k_x = \pm k_F/2$ there; this suggests effective barriers of $Z_{eff} = 0.5$ for the positive and $Z_{eff} = 1.5$ for the negative $k_x$ as indicated by the violet and orange colored dashed lines in Figs. 1(b)–(c).

### III. TUNNELING & TAHE CONDUCTANCES IN THE F REGION

Once all reflection coefficients are known, the tunneling (longitudinal) current in the F electrode can be obtained from the imbalance between the distributions of right-propagating and left-propagating charge carriers. The full calculation is rather technical and therefore, we only state the main points here. Throughout this work, we use the convention that positive current flows from left to right and negative current from right to left; the (positive) elementary charge is represented by $e$.

For an incident spin $\sigma$ electron from the left F electrode, the net resulting current flow is given by

$$\left(J_z^\sigma\right)^{\text{LEFT}} = e \sum_{\mathbf{k}_\parallel} \sum_{k_z^\sigma} \left\{ \left[ \frac{\hbar k_z^\sigma}{m} - \frac{\hbar k_z^\sigma}{m} \left|r_e^{\sigma,\sigma}(E)\right|^2 - \frac{\hbar k_z^{-\sigma}}{m} \left|r_e^{\sigma,-\sigma}(E)\right|^2 \right] f^0(E - eV) \left[1 - f^0(E)\right] \right.$$

$$\left. + \left[ -\frac{\hbar k_z^{-\sigma}}{m} \left|r_h^{\sigma,-\sigma}(E)\right|^2 - \frac{\hbar k_z^\sigma}{m} \left|r_h^{\sigma,\sigma}(E)\right|^2 \right] \left[1 - f^0(-E - eV)\right] f^0(-E) \right\}. \tag{S24}$$

The first summand refers to the incoming electron propagating from left to right with the positive (longitudinal) velocity $v_z^\sigma = (\hbar k_z^\sigma)/m$; the specularly and Andreev reflected electrons and holes propagate along the opposite direction, which we generally account for by using negative (longitudinal) velocities for the reflected parts. As the incident electron undergoes SR or AR only with certain probabilities, the respective current contributions need to be weighted by fractional parts which indicate that the corresponding scattering process occurs at all, i.e., by the related absolute squares of the reflection coefficients (which are themselves proportional to the *reflection probabilities*). Finally, appropriate products of Fermi-Dirac distribution functions [generally abbreviated with $f^0(E)$] are introduced to ensure that there are indeed occupied states for electrons (and likewise unoccupied ones for holes) in the left and unoccupied (occupied) states to tunnel into in the right electrode. Thereby, the longitudinal voltage drop $V$ along the junction has to be carefully included into the Fermi-Dirac functions.

In the same way, we can also compute the current associated with an incoming spin $\sigma$ electron(like quasiparticle) from the right electrode. Instead of dealing with quasiparticles in the S, it is more convenient to simply suppose an incident hole with spin $\sigma$ from the left (F) region; both descriptions are fully equivalent for computing transport properties. The related net current is found to read

$$\left(J_z^\sigma\right)^{\text{RIGHT}} = e \sum_{\mathbf{k}_\parallel} \sum_{k_z^\sigma} \left\{ \left[ -\frac{\hbar k_z^\sigma}{m} + \frac{\hbar k_z^\sigma}{m} \left|\tilde{r}_h^{\sigma,\sigma}(E)\right|^2 + \frac{\hbar k_z^{-\sigma}}{m} \left|\tilde{r}_h^{\sigma,-\sigma}(E)\right|^2 \right] \left[1 - f^0(E - eV)\right] f^0(E) \right.$$

$$\left. + \left[ \frac{\hbar k_z^{-\sigma}}{m} \left|\tilde{r}_e^{\sigma,-\sigma}(E)\right|^2 + \frac{\hbar k_z^\sigma}{m} \left|\tilde{r}_e^{\sigma,\sigma}(E)\right|^2 \right] f^0(-E - eV) \left[1 - f^0(-E)\right] \right\}. \tag{S25}$$

Note that the reflection coefficients for an incoming hole can in general be different from the ones we obtained for an incident electron in Sec. I; to stress this difference, we used "tilde reflection coefficients" in Eq. (S25). Particularly in the case of antisymmetric junctions, e.g., by having finite-size electrodes, the reflection probabilities for the different injection processes can indeed strongly differ and one should really proceed with the given general formulas when calculating the total current. However, owing to our junction's symmetry, we find that the total SR and AR probabilities averaged across the whole cross-section area are the same for incoming electrons and holes, i.e.,

$$\sum_{\mathbf{k}_\parallel} \left[ k_z^\sigma \left|r_e^{\sigma,\sigma}(E)\right|^2 + k_z^{-\sigma} \left|r_e^{\sigma,-\sigma}(E)\right|^2 \right] = \sum_{\mathbf{k}_\parallel} \left[ k_z^\sigma \left|\tilde{r}_h^{\sigma,\sigma}(E)\right|^2 + k_z^{-\sigma} \left|\tilde{r}_h^{\sigma,-\sigma}(E)\right|^2 \right] \tag{S26}$$

and

$$\sum_{\mathbf{k}_\parallel} \left[ k_z^{-\sigma} \left|r_h^{\sigma,-\sigma}(E)\right|^2 + k_z^\sigma \left|r_h^{\sigma,\sigma}(E)\right|^2 \right] = \sum_{\mathbf{k}_\parallel} \left[ k_z^{-\sigma} \left|\tilde{r}_e^{\sigma,-\sigma}(E)\right|^2 + k_z^\sigma \left|\tilde{r}_e^{\sigma,\sigma}(E)\right|^2 \right]. \tag{S27}$$

This allows us to obtain a rather simple expression for the total current by combining Eqs. (S24) and (S25), rewriting the sums over $\mathbf{k}_\parallel$ and $k_z^\sigma$ as integrals within continuum approximation ($A$ is the interfacial cross-section area), and finally summing over

both spin channels,

$$J_z = \sum_{\sigma=\pm 1} \left[(J_z^\sigma)^{\text{LEFT}} + (J_z^\sigma)^{\text{RIGHT}}\right]$$

$$= e \sum_{\sigma=\pm 1} \frac{A}{(2\pi)^3} \int d^2\mathbf{k}_\parallel \int_{-\infty}^{\infty} dk_z^\sigma \left\{\left[\frac{\hbar k_z^\sigma}{m} - \frac{\hbar k_z^\sigma}{m} \left|r_e^{\sigma,\sigma}(E)\right|^2 - \frac{\hbar k_z^{-\sigma}}{m} \left|r_e^{\sigma,-\sigma}(E)\right|^2\right] \left[f^0(E - eV) - f^0(E)\right]\right.$$

$$\left. + \left[-\frac{\hbar k_z^{-\sigma}}{m} \left|r_h^{\sigma,-\sigma}(E)\right|^2 - \frac{\hbar k_z^\sigma}{m} \left|r_h^{\sigma,\sigma}(E)\right|^2\right] \left[f^0(E + eV) - f^0(E)\right]\right\}$$

$$= e \sum_{\sigma=\pm 1} \frac{A}{(2\pi)^3} \int d^2\mathbf{k}_\parallel \int_{-\infty}^{\infty} dE \left(\frac{\partial k_z^\sigma}{\partial E}\right) \left\{\left[\frac{\hbar k_z^\sigma}{m} - \frac{\hbar k_z^\sigma}{m} \left|r_e^{\sigma,\sigma}(E)\right|^2 - \frac{\hbar k_z^{-\sigma}}{m} \left|r_e^{\sigma,-\sigma}(E)\right|^2\right] \left[f^0(E - eV) - f^0(E)\right]\right.$$

$$\left. + \left[-\frac{\hbar k_z^{-\sigma}}{m} \left|r_h^{\sigma,-\sigma}(E)\right|^2 - \frac{\hbar k_z^\sigma}{m} \left|r_h^{\sigma,\sigma}(E)\right|^2\right] \left[f^0(E + eV) - f^0(E)\right]\right\}. \quad (S28)$$

Equation (S28) is valid at arbitrary temperatures through the general expressions for the Fermi-Dirac functions. The junction's differential tunneling conductance is then obtained from $G_{z,z} = dJ_z/dV$; particularly at zero temperature, we may use the relation $df^0(E)/dE \approx -\delta(E)$ to obtain the compact expression

$$G_{z,z} = \frac{G_0 A}{2(2\pi)^2} \sum_{\sigma=\pm 1} \int d^2\mathbf{k}_\parallel \left\{1 - \left[\left|r_e^{\sigma,\sigma}(eV)\right|^2 + \frac{k_z^{-\sigma}}{k_z^\sigma} \left|r_e^{\sigma,-\sigma}(eV)\right|^2\right] + \left[\frac{k_z^{-\sigma}}{k_z^\sigma} \left|r_h^{\sigma,-\sigma}(-eV)\right|^2 + \left|r_h^{\sigma,\sigma}(-eV)\right|^2\right]\right\}, \quad (S29)$$

with the usual conductance quantum $G_0 = (2e^2)/h$. Equation (S29) is the generalization of the BTK conductance formula [S11] to the case of F/SC/S junctions with interfacial SOC [S12].

To derive the transverse currents, $J_\eta$ ($\eta \in \{x; y\}$), and the related transverse TAHE conductances, $G_{\eta,z}$, we follow a similar procedure. The most crucial point in the derivation is to properly replace the velocities along the $\hat{z}$-direction by those along the $\hat{\eta}$-direction without messing up the signs. Taking into account the contributions of an incident spin $\sigma$ electron from the F and another one from the S (which is again equivalent to having an incoming hole from the F) yields for the transverse currents

$$J_\eta = e \sum_{\sigma=\pm 1} \sum_{\mathbf{k}_\parallel} \sum_{k_z^\sigma} \left\{\left[\frac{\hbar k_\eta}{m} - \frac{\hbar k_\eta}{m} \left|r_e^{\sigma,\sigma}(E)\right|^2 - \frac{\hbar k_\eta}{m} \left|r_e^{\sigma,-\sigma}(E)\right|^2\right] f^0(E - eV) \left[1 - f^0(E)\right]\right.$$

$$+ \left[\frac{\hbar k_\eta}{m} \left|r_h^{\sigma,-\sigma}(E)\right|^2 + \frac{\hbar k_\eta}{m} \left|r_h^{\sigma,\sigma}(E)\right|^2\right] \left[1 - f^0(-E - eV)\right] f^0(-E)$$

$$+ \left[-\frac{\hbar k_\eta}{m} - \frac{\hbar k_\eta}{m} \left|\tilde{r}_h^{\sigma,\sigma}(E)\right|^2 - \frac{\hbar k_\eta}{m} \left|\tilde{r}_h^{\sigma,-\sigma}(E)\right|^2\right] \left[1 - f^0(E - eV)\right] f^0(E)$$

$$\left. + \left[\frac{\hbar k_\eta}{m} \left|\tilde{r}_e^{\sigma,-\sigma}(E)\right|^2 + \frac{\hbar k_\eta}{m} \left|\tilde{r}_e^{\sigma,\sigma}(E)\right|^2\right] f^0(-E - eV) \left[1 - f^0(-E)\right]\right\}. \quad (S30)$$

The physical meaning of the different contributions is illustrated and explained in Fig. S1 and its caption, and therefore, we do not discuss it here.

As before, we can express the SR and AR coefficients in the case of an incident hole via the known ones for an incoming electron by exploiting the symmetries in our system; specifically, we unravel

$$\sum_{\mathbf{k}_\parallel} \left[k_\eta \left|r_e^{\sigma,\sigma}(E)\right|^2 + k_\eta \left|r_e^{\sigma,-\sigma}(E)\right|^2\right] = -\sum_{\mathbf{k}_\parallel} \left[k_\eta \left|\tilde{r}_h^{\sigma,\sigma}(E)\right|^2 + k_\eta \left|\tilde{r}_h^{\sigma,-\sigma}(E)\right|^2\right] \quad (S31)$$

and

$$\sum_{\mathbf{k}_\parallel} \left[k_\eta \left|r_h^{\sigma,-\sigma}(E)\right|^2 + k_\eta \left|r_h^{\sigma,\sigma}(E)\right|^2\right] = -\sum_{\mathbf{k}_\parallel} \left[k_\eta \left|\tilde{r}_e^{\sigma,-\sigma}(E)\right|^2 + k_\eta \left|\tilde{r}_e^{\sigma,\sigma}(E)\right|^2\right], \quad (S32)$$

which simplifies the total transverse currents to

$$J_\eta = e \sum_{\sigma=\pm 1} \frac{A}{(2\pi)^3} \int d^2\mathbf{k}_\parallel \int_{-\infty}^{\infty} dE \left(\frac{\partial k_z^\sigma}{\partial E}\right) \left\{\left[\frac{\hbar k_\eta}{m} - \frac{\hbar k_\eta}{m} \left|r_e^{\sigma,\sigma}(E)\right|^2 - \frac{\hbar k_\eta}{m} \left|r_e^{\sigma,-\sigma}(E)\right|^2\right] \left[f^0(E - eV) - f^0(E)\right]\right.$$

$$\left. + \left[\frac{\hbar k_\eta}{m} \left|r_h^{\sigma,-\sigma}(E)\right|^2 + \frac{\hbar k_\eta}{m} \left|r_h^{\sigma,\sigma}(E)\right|^2\right] \left[f^0(E + eV) - f^0(E)\right]\right\}. \quad (S33)$$

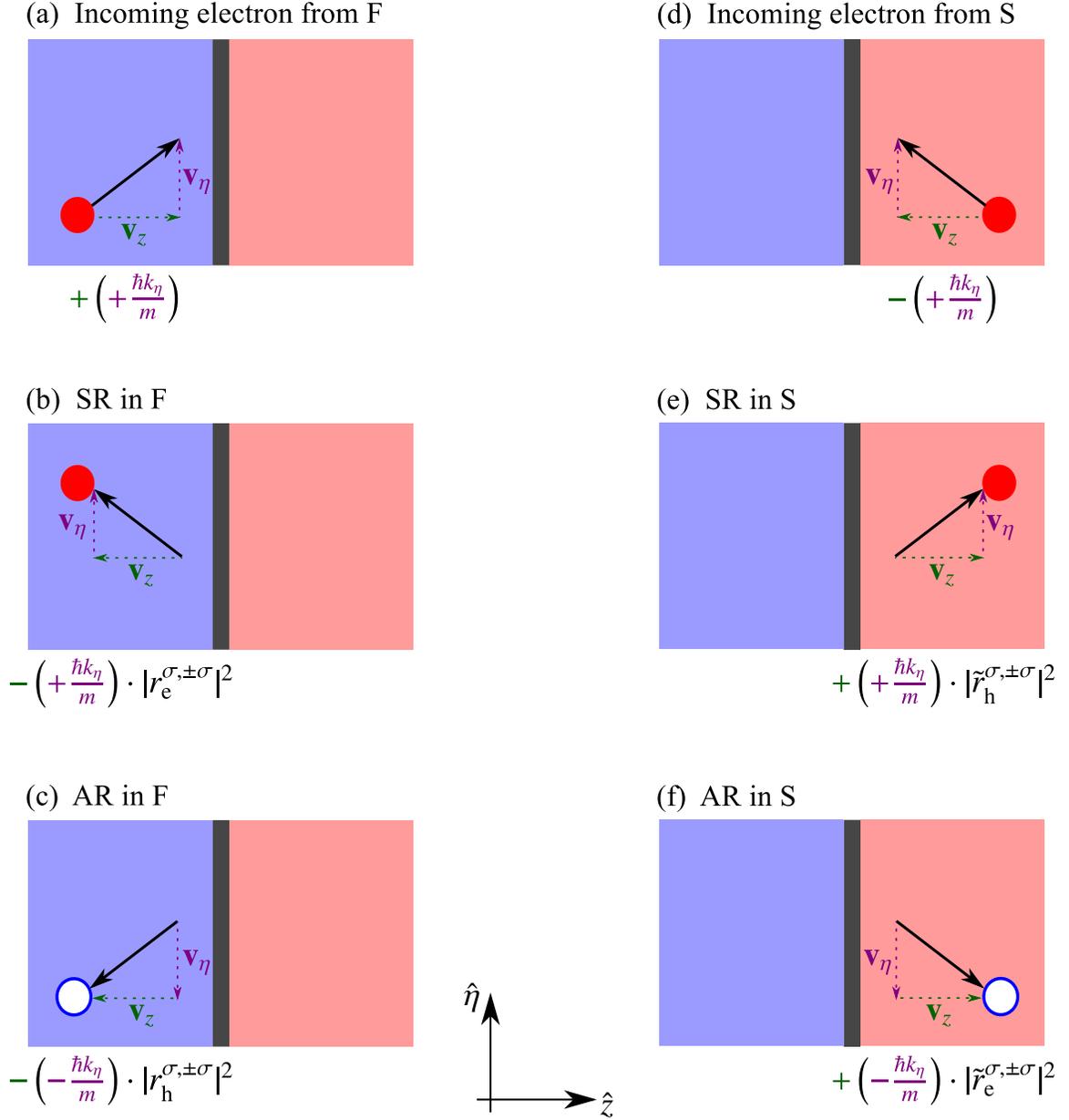

FIG. S1. Illustration of the different scattering processes' contributions to the full transverse current flow computed by means of Eq. (S30). The panels (a)–(f) reflect the contributions in the same order as they enter Eq. (S30). Panel (a) refers to the incident spin $\sigma$ electron from the F; its transverse velocity along the $\hat{\eta}$-direction is $|\mathbf{v}_\eta| = (+\hbar k_\eta)/m$ (violet) and, since the electron propagates towards the barrier, we count its net contribution to the transverse current positively (green positive sign). Panels (b) and (c) indicate the two possible reflection processes of the incoming electron: SR and AR. In the first case, the reflected electron's transverse velocity does not change, while in the second scenario, the reflected hole has opposite transverse velocity. In contrast to the incoming electron, the reflected particles propagate away from the barrier (along $-\hat{z}$) and thus, in order to get a balanced current flow in the barrier region, we have to subtract their contributions from the incident current (green negative signs). Both processes only occur with certain probabilities, which we account for by multiplying the contributions by the reflection coefficients' absolute squares. Panels (d)–(f) show analogous considerations for an incoming electron(like quasiparticle) from the S. As mentioned in the text, instead of dealing with quasiparticles in the S, it is more convenient to suppose an incident spin $\sigma$ hole from the F; we followed that approach in Eq. (S30), which particularly means that processes (e) and (f) gain an additional negative sign in Eq. (S30) when compared to the depicted situation (as we need to transfer the shown processes' scattering coefficients from the electron- into the hole-picture). This is also the deeper reason behind the notation of the "tilde reflection coefficients". Finally, all described parts must be properly multiplied by Fermi-Dirac distribution functions in Eq. (S30), ensuring that the incoming and scattered states are indeed occupied and unoccupied as the different scattering processes require. Note that there are several other ways to construct an ansatz for the transverse current, which are all equivalent to ours and lead to the same final result [see Eq. (S34)] in the end.

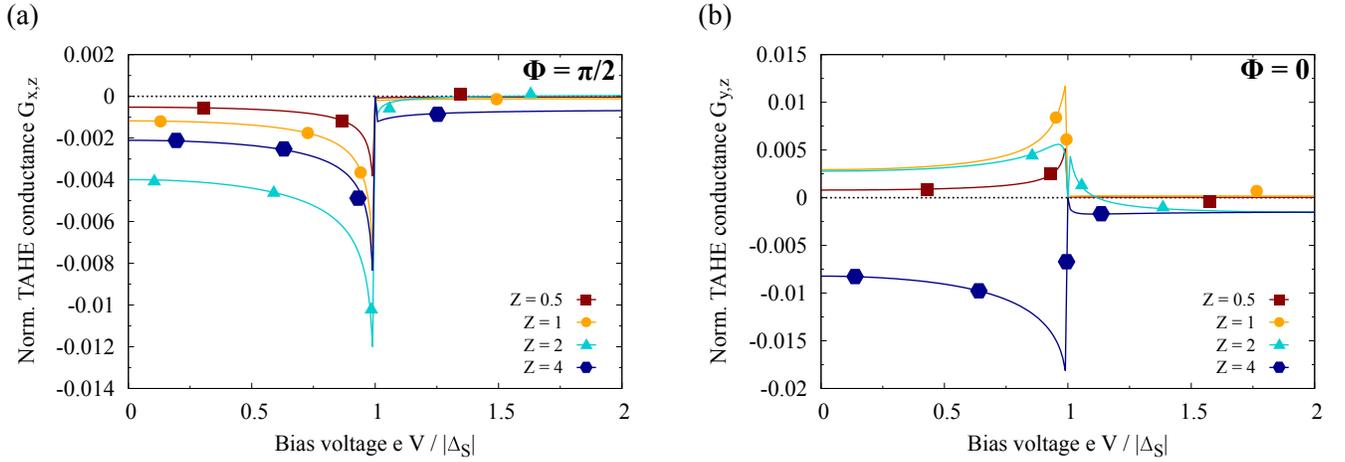

FIG. S2. Calculated dependence of the maximal TAHE conductances, (a) $G_{x,z}(\Phi = \pi/2)$ and $G_{y,z}(\Phi = 0)$, normalized to Sharvin's conductance, on the applied bias voltage $V$, and for various indicated barrier strengths $Z$; the SOC parameters are $\alpha = 42.3\,\text{eV}\,\text{Å}^2$ and $\beta \approx 19.2\,\text{eV}\,\text{Å}^2 \cdot Z$.

At zero temperature, the transverse TAHE conductance components, $G_{\eta,z} = dJ_\eta/dz$, can again be approximated as

$$G_{\eta,z} = \frac{G_0 A}{2(2\pi)^2} \sum_{\sigma=\pm 1} \int d^2\mathbf{k}_\| \frac{k_\eta}{k_z^\sigma} \left\{ 1 - \left[ |r_e^{\sigma,\sigma}(eV)|^2 + |r_e^{\sigma,-\sigma}(eV)|^2 \right] + \left[ |r_h^{\sigma,-\sigma}(-eV)|^2 + |r_h^{\sigma,\sigma}(-eV)|^2 \right] \right\}$$

$$= -\frac{G_0 A}{2(2\pi)^2} \sum_{\sigma=\pm 1} \int d^2\mathbf{k}_\| \frac{k_\eta}{k_z^\sigma} \left\{ \left[ |r_e^{\sigma,\sigma}(eV)|^2 + |r_e^{\sigma,-\sigma}(eV)|^2 \right] + \left[ |r_h^{\sigma,-\sigma}(-eV)|^2 + |r_h^{\sigma,\sigma}(-eV)|^2 \right] \right\}; \quad \text{(S34)}$$

in the last step, we used $\int d^2\mathbf{k}_\| k_\eta/k_z^\sigma = 0$ due to symmetry. The final result is given as Eq. (3) in the manuscript. As already mentioned there, the SR and AR parts remarkably enter with the same sign into the final expression for the transverse currents, contrarily to what has been established in the BTK model for the tunneling (longitudinal) current. This observation can be physically justified since—in contrast to the longitudinal velocity components, which point into the same direction—the specularly reflected electron's and the Andreev reflected hole's transverse velocity vectors point along opposite directions (see Fig. S1); the accumulated relative sign between both contributions is then exactly compensated by the fact that holes possess opposite charge. Finally, the SR and AR parts must appear with same signs in the final transverse current formula. This confirms our statement made in the manuscript when discussing the skew reflection mechanism: skew SR and AR can indeed constructively amplify the TAHE, thus noticeably enhancing the associated conductances compared to the normal-state counterpart.

For the data presented in Figs. 2 and 3 of the manuscript (zero-bias scenario), we evaluated Eq. (S34) and the analytically obtained reflection coefficients [see Eqs. (S10)–(S14) in the superconducting and Eqs. (S15)–(S19) in the normal-conducting state]; the integration over $\mathbf{k}_\|$ was performed numerically by means of a two-dimensional Gaussian quadrature algorithm, where the upper integration limit in the $|\mathbf{k}_\||$-integration was adapted such that we ensure propagating states for the incident electrons, $|\mathbf{k}_\||_{\text{max}}^\sigma = \min[k_F \sqrt{1 + \sigma P}; k_F]$. At finite bias voltages [see Figs. S2–S3], we additionally obtained the reflection coefficients numerically from the generic system of equations following from the methodology described in Sec. I and then proceeded with the general equations for the conductance components.

## IV. SYMMETRY ANALYSIS OF THE TAHE CONDUCTANCES

Given the general form of the TAHE conductances, Eq. (S34), we may have a closer look at the underlying symmetries to extract valuable information about the conductances' characteristic scaling behavior, for instance, with respect to the magnetization angle $\Phi$; see also Ref. [S13]. The latter quantity's impact is hidden in the general form of the reflection coefficients; see, e.g., Eqs. (S10)–(S13). For a fixed $\mathbf{k}_\|$, the system has only two preferred spatial directions: one oriented along the spin-orbit fields, $\mathbf{\Omega}(\mathbf{k}_\|)$, and another one along the magnetization unit vector, $\hat{\mathbf{m}}$. Exploiting this observation, we may expand the reflection coefficient-dependent part in Eq. (S34) for weak SOC by means of a Taylor series in $[\hat{\mathbf{m}} \cdot \mathbf{\Omega}(\mathbf{k}_\|)]$, which has the general form

$$-\left\{ \left[ |r_e^{\sigma,\sigma}(eV)|^2 + |r_e^{\sigma,-\sigma}(eV)|^2 \right] + \left[ |r_h^{\sigma,-\sigma}(-eV)|^2 + |r_h^{\sigma,\sigma}(-eV)|^2 \right] \right\} \approx \sum_{i=0}^{\infty} \xi^{(i)} \left[ \hat{\mathbf{m}} \cdot \mathbf{\Omega}(\mathbf{k}_\|) \right]^{(i)}. \quad \text{(S35)}$$

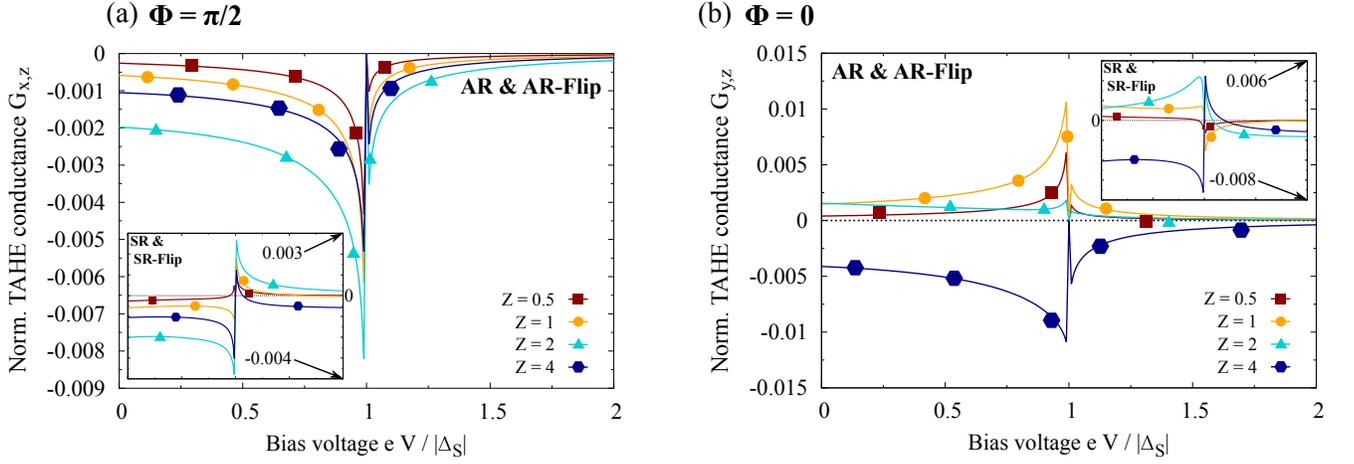

FIG. S3. Calculated dependence of the maximal TAHE conductances, (a) $G_{x,z}(\Phi = \pi/2)$ and (b) $G_{y,z}(\Phi = 0)$, normalized to Sharvin's conductance, on the applied bias voltage $V$ and for various indicated barrier strengths $Z$; the SOC parameters are $\alpha = 42.3\,\text{eV}\,\text{Å}^2$ and $\beta \approx 19.2\,\text{eV}\,\text{Å}^2 \cdot Z$. The plots show the conductance contributions stemming from spin-conserving and spin-flip AR; the contributions of the corresponding SR processes are depicted in the insets.

It is not necessary for the following analysis to specify the concrete form of the expansion coefficients $\xi^{(i)}$; the only important observation is that as SOC is assumed to be weak, these expansion coefficients can be evaluated when SOC is absent [Eq. (S35) then treats the situation with nonzero SOC in a perturbative-like way]. Inserting Eq. (S35) into Eq. (S34) and terminating the Taylor series after the second order yields for the approximated transverse conductances (in the presence of SOC)

$$G_{\eta,z} \approx G_{\eta,z}^{(0)} + G_{\eta,z}^{(1)} + G_{\eta,z}^{(2)} + \ldots, \tag{S36}$$

with the single constituents

$$G_{\eta,z}^{(i)} = \frac{G_0 A}{2(2\pi)^2} \sum_{\sigma=\pm 1} \int d^2\mathbf{k}_\parallel \frac{k_\eta}{k_z^\sigma} \xi^{(i)} \left[\hat{\mathbf{m}} \cdot \boldsymbol{\Omega}(\mathbf{k}_\parallel)\right]^{(i)}. \tag{S37}$$

Let us at first focus for a moment on $G_{x,z}$. Substituting polar coordinates, $k_x = |\mathbf{k}_\parallel| \cos\theta$ and $k_y = |\mathbf{k}_\parallel| \sin\theta$, and writing out the two-dimensional integral results in

$$G_{x,z}^{(i)} = \frac{G_0 A}{2(2\pi)^2} \sum_{\sigma=\pm 1} \int_0^{|\mathbf{k}_\parallel|_{\text{max}}^\sigma} d|\mathbf{k}_\parallel| \int_0^{2\pi} d\theta \frac{|\mathbf{k}_\parallel|\cos\theta}{k_z^\sigma} \xi^{(i)} \left[\hat{\mathbf{m}} \cdot \boldsymbol{\Omega}(\mathbf{k}_\parallel)\right]^{(i)}. \tag{S38}$$

By inspecting the different factors' parities with respect to changing the sign of $\mathbf{k}_\parallel$, we can immediately deduce which of the integrals finally entering Eq. (S36) leads to nonzero contributions (to lowest order). There are only two parts with *odd* $\mathbf{k}_\parallel$-parity: $k_x$ and $\hat{\mathbf{m}} \cdot \boldsymbol{\Omega}$ [since $\boldsymbol{\Omega}(-\mathbf{k}_\parallel) = -\boldsymbol{\Omega}(\mathbf{k}_\parallel)$]. As a consequence, only those $G_{\eta,z}^{(i)}$'s in Eq. (S36) containing odd powers of $\hat{\mathbf{m}} \cdot \boldsymbol{\Omega}$ can give rise to a totally *even* $\mathbf{k}_\parallel$-integrand and will hence not vanish after performing the $\mathbf{k}_\parallel$-integration over the whole two-dimensional space parallel to the interface. In our case, only $G_{x,z}^{(1)}$ fulfills all these requirements to lead to a nonzero conductance contribution.

After plugging in the explicit form of $\boldsymbol{\Omega}$, we obtain in polar coordinates

$$G_{x,z} \approx G_{x,z}^{(1)} \sim \sum_{\sigma=\pm 1} \int_0^{|\mathbf{k}_\parallel|_{\text{max}}^\sigma} d|\mathbf{k}_\parallel| \int_0^{2\pi} d\theta \left[(\alpha - \beta)\cos\Phi |\mathbf{k}_\parallel|^2 \cos\theta \sin\theta - (\alpha + \beta)\sin\Phi |\mathbf{k}_\parallel|^2 \cos^2\theta\right]$$

$$\sim -(\alpha + \beta)\sin\Phi. \tag{S39}$$

In the last step, we again relied on our parity argumentation: the integrand's first summand has *odd* $\mathbf{k}_\parallel$-parity and vanishes after performing the $\mathbf{k}_\parallel$-integration. The characteristic scaling behavior of $G_{x,z}$ with respect to $\Phi$, $G_{x,z}(\Phi) \sim -(\alpha + \beta)\sin\Phi$, solely follows from symmetry arguments and is a characteristic fingerprint of the TAHE conductance as stated in the manuscript; it has already been found in the normal-conducting TAHE [S13, S14] and remains unaffected in superconducting systems. An analogous treatment of $G_{y,z}$ suggests $G_{y,z}(\Phi) \sim (\alpha - \beta)\cos\Phi$. Therefore, the interference of simultaneously present Rashba and Dresselhaus SOC leads to a clearly pronounced magnetoanisotropy between $G_{x,z}$ and $G_{y,z}$ as stated in the manuscript; the maximal conductance amplitudes scale as a function of the SOC strengths as $G_{x,z}(\Phi = \pi/2) \sim -(\alpha + \beta)$ and $G_{y,z}(\Phi = 0) \sim \alpha - \beta$,

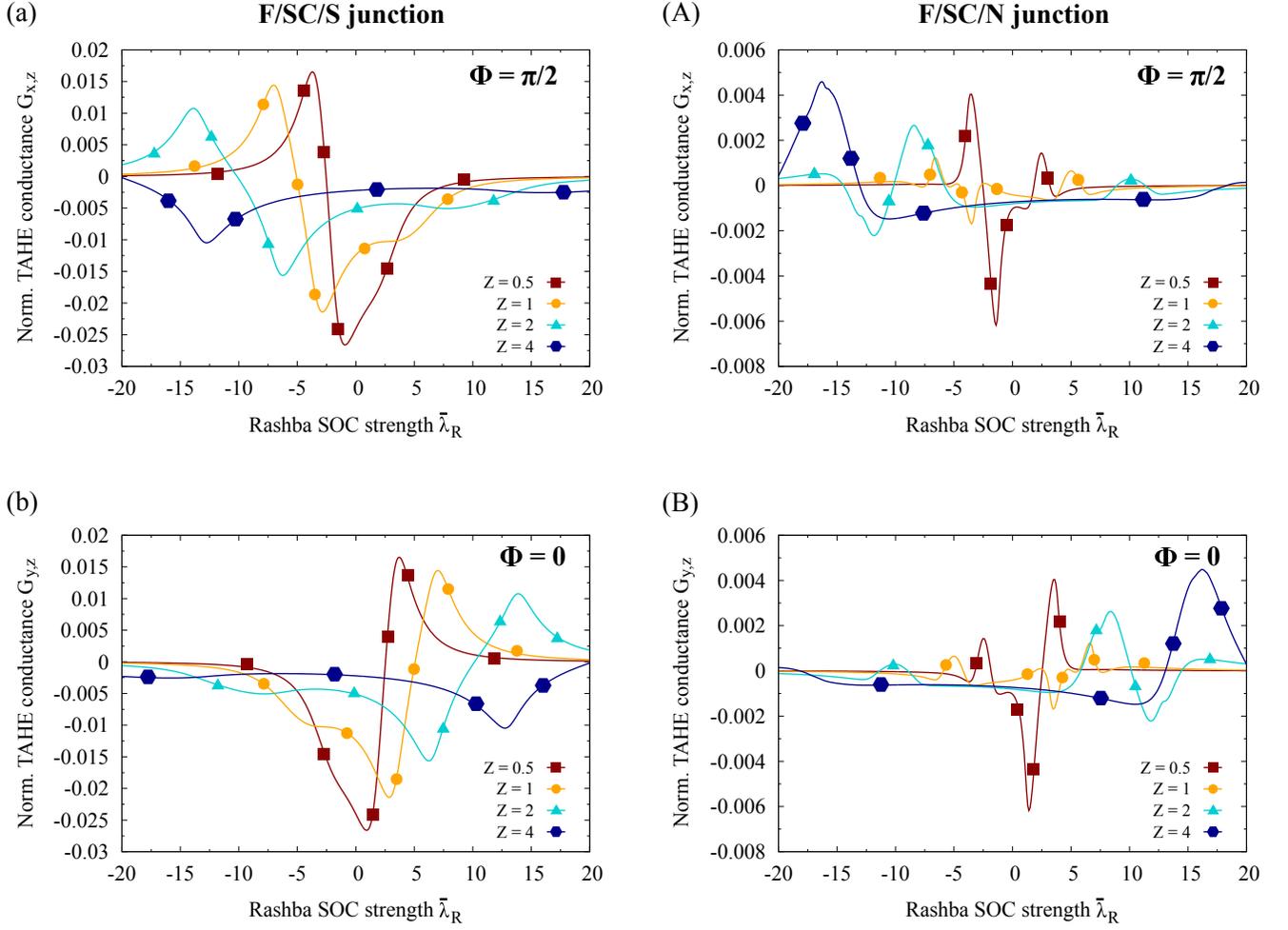

FIG. S4. Calculated dependence of the maximal zero-bias TAHE conductances, (a) $G_{x,z}(\Phi = \pi/2)$ and (b) $G_{y,z}(\Phi = 0)$, normalized to Sharvin's conductance, on the dimensionless Rashba SOC parameter $\bar{\lambda}_R = (2m\alpha)/\hbar^2$ ($\alpha = 42.3\,\text{eV\,Å}^2$ refers to $\bar{\lambda}_R \sim 11.2$), and for various indicated barrier strengths $Z$. The Dresselhaus SOC parameter is $\beta \approx 19.2\,\text{eV\,Å}^2 \cdot Z$. (A) and (B) show the corresponding normal-state calculations.

respectively. If Dresselhaus SOC is absent and only Rashba SOC remains finite, the anisotropy between $G_{x,z}$ and $G_{y,z}$ disappears and the maximal amplitudes become equal (up to a relative sign). All our analyzed (numerical) calculations reflect exactly the expected symmetries.

## V. TRANSVERSE SUPERCURRENT RESPONSE IN THE S REGION

We stated in the manuscript that interpreting the existence of TAHEs in magnetic tunnel junctions in terms of momentum- and spin-dependent skew AR and SR of particles at the SC interface suggests that also the Cooper pairs transferred into the superconducting electrode of our junction via the AR processes must encounter a similar effect, eventually leading to a spontaneously generated transverse supercurrent response there. To compute transverse supercurrents in the S layer (for simplicity, at zero external bias), we follow the well-established Green's function-based technique, which was firstly introduced by Furusaki and Tsukada [S15] to compute Josephson current flows. Since both electrodes of our system span semi-infinite regions, it does actually not matter for transport calculations which of them is located at $z < 0$ and which at $z > 0$. To avoid too many confusing negative signs in the propagating wave functions and to be consistent with the notation introduced in Ref. [S15], we assume for a moment that the S electrode is at $z < 0$ and the F at $z > 0$, respectively. The BdG scattering states for an incoming electronlike

quasiparticle with spin up (spin parallel to the F's magnetization) can then be written as

$$\psi^{(1)}(z<0) = \begin{bmatrix} u \\ 0 \\ v \\ 0 \end{bmatrix} e^{iq_z z} + \mathcal{A}^{(1)} \begin{bmatrix} u \\ 0 \\ v \\ 0 \end{bmatrix} e^{-iq_z z} + \mathcal{B}^{(1)} \begin{bmatrix} 0 \\ u \\ 0 \\ v \end{bmatrix} e^{-iq_z z} + C^{(1)} \begin{bmatrix} v \\ 0 \\ u \\ 0 \end{bmatrix} e^{iq_z z} + \mathcal{D}^{(1)} \begin{bmatrix} 0 \\ v \\ 0 \\ u \end{bmatrix} e^{iq_z z} \tag{S40}$$

and

$$\psi^{(1)}(z>0) = \mathcal{E}^{(1)} \frac{1}{\sqrt{2}} \begin{bmatrix} 1 \\ e^{i\Phi} \\ 0 \\ 0 \end{bmatrix} e^{ik_z^{\sigma=1} z} + \mathcal{F}^{(1)} \frac{1}{\sqrt{2}} \begin{bmatrix} 1 \\ -e^{i\Phi} \\ 0 \\ 0 \end{bmatrix} e^{ik_z^{\sigma=-1} z}$$

$$+ \mathcal{G}^{(1)} \frac{1}{\sqrt{2}} \begin{bmatrix} 0 \\ 0 \\ 1 \\ -e^{i\Phi} \end{bmatrix} e^{-ik_z^{\sigma=1} z} + \mathcal{H}^{(1)} \frac{1}{\sqrt{2}} \begin{bmatrix} 0 \\ 0 \\ 1 \\ e^{i\Phi} \end{bmatrix} e^{-ik_z^{\sigma=-1} z}; \tag{S41}$$

the BCS coherence factors, as well as the wave vectors, are the same as declared in Sec. I. The calligraphically written scattering coefficients are again obtained by applying appropriate boundary conditions, similarly to Eqs. (S8)–(S9), to the states and solving the resulting system of equations. Particularly important is the spin up AR coefficient $C^{(1)}$. Similar considerations are repeated for an incident electronlike quasiparticle with spin down and holelike quasiparticles; the corresponding AR coefficients required for calculating the supercurrent are in those situations denoted by $\mathcal{D}^{(2)}$, $\mathcal{A}^{(3)}$, and $\mathcal{B}^{(4)}$, accordingly. Once all four AR coefficients are determined, the transverse supercurrent components (including singlet and triplet contributions *simultaneously*) in the S layer close to the interfacial SC, $I_\eta$ ($\eta \in \{x, y\}$), are obtained from

$$I_\eta = \frac{e k_B T}{2\hbar} |\Delta_S| \sum_{\mathbf{k}_\parallel} \sum_{\omega_n} \frac{k_\eta}{\sqrt{k_F^2 - \mathbf{k}_\parallel^2}} \left[ \frac{C^{(1)}(i\omega_n) + \mathcal{D}^{(2)}(i\omega_n) + \mathcal{A}^{(3)}(i\omega_n) + \mathcal{B}^{(4)}(i\omega_n)}{\sqrt{\omega_n^2 + |\Delta_S|^2}} \right]. \tag{S42}$$

Equation (S42) can be derived from the system's Matsubara Green's function [S8, S15, S16]; $\omega_n = (2n + 1)\pi k_B T$ with integer $n$ denotes the (fermionic) Matsubara frequencies, while $k_B$ stands for Boltzmann's constant. Although the thermal energy, $k_B T$, appears in the numerator of Eq. (S42), one should be aware that also the AR coefficients depend on temperature in a nontrivial way. Therefore, although it might not be too obvious at a first glance, Eq. (S42) is also well-applicable at zero temperature by properly considering the limit $T \to 0_+$. For Fig. 4 in the manuscript, we evaluated Eq. (S42) fully numerically for the indicated parameters.

## VI. DEPENDENCE OF TAHE CONDUCTANCES ON BIAS VOLTAGE

In the manuscript, we analyzed the TAHE conductances at zero external bias voltage. Nevertheless, our presented analytical approach is general enough to treat also the situation of finite bias. Figure S2 illustrates the scaling of the maximal TAHE conductances, (a) $G_{x,z}(\Phi = \pi/2)$ and (b) $G_{y,z}(\Phi = 0)$, with increasing bias voltage $V$ ($e$ and $|\Delta_S|$ are kept constant). We particularly focus on the subgap regime $eV/|\Delta_S| \leq 1$ that corresponds to superconducting junctions; considering $eV/|\Delta_S| \gg 1$ effectively reduces our model simply to the description of its normal-conducting counterpart. As explained in the manuscript (at zero bias voltage), the TAHE conductances' amplitudes become sizable in the superconducting and dramatically damped in the normal-conducting state. Increasing the external bias voltage in superconducting junctions even results in a further remarkable increase of the TAHE conductances (without changing the relative orientations of the Hall currents), reaching a clearly pronounced transverse conductance peak slightly below the superconducting gap. Exactly at the gap edge, $eV/|\Delta_S| = 1$, both TAHE conductances drop exactly to zero before finally approaching the small (but finite) values indicating the normal-state TAHE analog. All the observed characteristics can again be traced back to the effective skew reflection model introduced in the manuscript; see Fig. 1 and the related explanations there. At finite bias, the AR and SR probabilities modulate much faster with a change of the effective scattering potential strength than at zero voltage. As a consequence, the asymmetry between the scattering potentials incident particles with positive and negative transverse momentum are exposed to is much more pronounced; that creates a much stronger spatial charge imbalance and larger Hall current flows (TAHE conductances).

Particularly at the gap edge, $eV/|\Delta_S| = 1$, both the AR and SR probabilities approach constant values and get completely independent of the effective interfacial scattering potential. In the ideal case of a N/S junction, BTK [S11] predicts perfect AR at $eV = |\Delta_S|$ (SR and quasiparticle transmissions into the S are forbidden), finally leading to the formation of a clear finite-bias

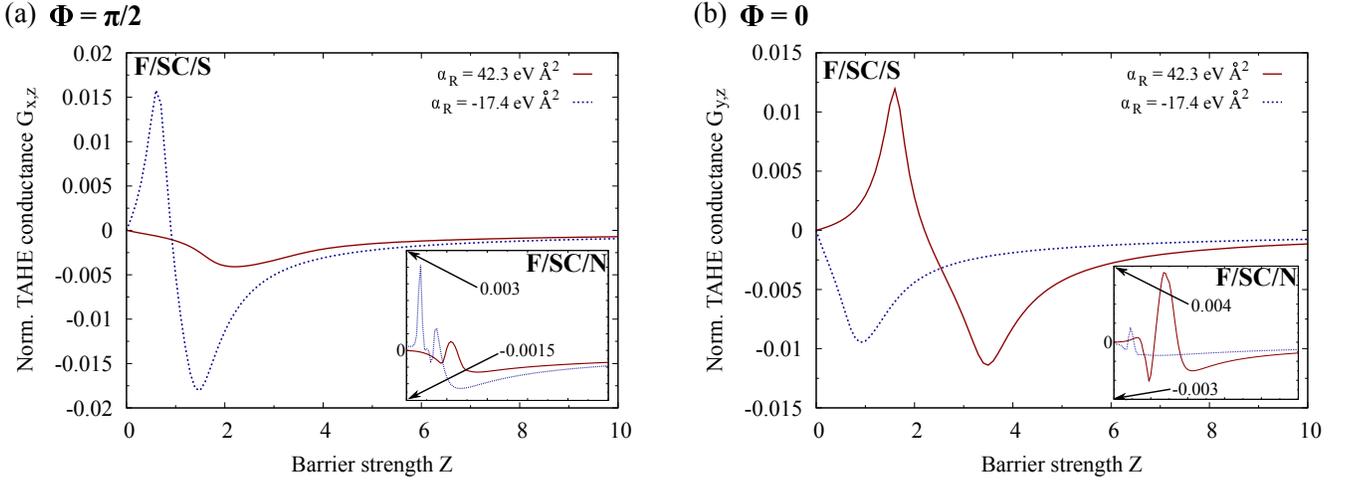

FIG. S5. Calculated dependence of the maximal zero-bias TAHE conductances, (a) $G_{x,z}(\Phi = \pi/2)$ and (b) $G_{y,z}(\Phi = 0)$, normalized to Sharvin's conductance, on the barrier strength $Z$ for two indicated Rashba SOC parameters, $\alpha = 42.3\,\text{eV}\,\text{Å}^2$ and $\alpha = -17.4\,\text{eV}\,\text{Å}^2$; the Dresselhaus SOC parameter is $\beta \approx 19.2\,\text{eV}\,\text{Å}^2 \cdot Z$. The insets show the corresponding normal-state calculations.

conductance peak. In our situation, the AR probability is slightly damped by the presence of the finite exchange splitting in the F that gives also rise to a small probability for SR. Nevertheless, both probabilities are still fully independent of the scattering potential so that the skew reflection mechanism cannot produce any spatial charge asymmetry and no Hall current can start to flow (that means zero TAHE conductance as calculated). The latter finding might play an important role to experimentally distinguish the underlying physical mechanisms of the Hall effect; the Hall conductances drop only to exact zero at $eV = |\Delta_S|$ in the case of the TAHE due to the described peculiarities of the skew reflection picture, whereas possibly additionally present conventional anomalous Hall effects give rise to finite Hall conductances even at the gap edge.

Similarly to our analysis in the manuscript, we illustrate in Fig. S3 the single contributions to the TAHE conductances that can be associated to the various appearing skew reflection processes at the interface. Shortly speaking, the results confirm the previously stated qualitative trends. While skew AR and SR parts are quantitatively equally important to generate sizable TAHE conductances *at zero bias* (they indeed act together and enhance the effect there), we see by comparing the resolved calculations to the full ones in Fig. S2 that particularly the qualitative variations occurring at finite bias are mostly governed by skew AR. Moreover, the pronounced TAHE conductance peaks close to the gap edge stem also predominantly from skew AR as one can most clearly see in the data belonging to $G_{x,z}$. We could have expected that from our argumentation in the preceding paragraph, suggesting that *AR is the dominant reflection process close to the gap edge*. Therefore, the TAHE conductances must again vanish exactly at the gap edge; see our explanation above.

## VII. ROLE OF RASHBA SOC, BARRIER STRENGTH, AND THE F'S SPIN POLARIZATION

Up to now, we have not answered the questions how changing the strength of the interfacial Rashba SOC (Dresselhaus SOC is determined by the material) by electrical gating [S17–S21] or designing different interface geometries, altering the barrier's strength (height and width), or replacing the F by one with different spin polarization might impact the investigated TAHEs.

To address the first question, Fig. S4 shows the maximal (zero-bias) TAHE conductances for various barrier strengths $Z$ and as a function of the effective dimensionless Rashba SOC parameter $\bar{\lambda}_R = (2m\alpha)/\hbar^2$; $\alpha = 42.3\,\text{eV}\,\text{Å}^2$ chosen for all calculations in the manuscript refers to $\bar{\lambda}_R \sim 11.2$. Both $G_{x,z}$ and $G_{y,z}$ modulate in a highly nonmonotonic way when changing $\bar{\lambda}_R$ or $Z$. As a consequence, there are in principle two practical ways to reverse the orientation of the flowing transverse Hall currents: either by increasing the barrier strength at certain $\bar{\lambda}_R$'s (as we observed in the $G_{y,z}$–$\Phi$ relations presented in Fig. 2 of the manuscript) or by tuning $\bar{\lambda}_R$ for fixed $Z$ (see also the next paragraph). The physical reason behind this must be attributed to the specific form of the effective interfacial scattering potential, $V_\text{eff}$ [see Eq. (2) in the manuscript], which couples the (usual) barrier strength with the SOC-dependent skew reflection part. Therefore, appropriately chosen combinations of $Z$ and $\bar{\lambda}_R$ can reverse the $V_\text{eff}$-dependent asymmetry in the reflection probabilities, e.g., by creating a situation in which the particle with positive $k_x$ now formally feels a strong "negative" $V_\text{eff}$, while the particle with negative $k_x$ resides at a weak positive $V_\text{eff}$. Given that the BTK-like reflection probabilities are symmetric under a sign change of $V_\text{eff}$, we conclude that now the particle with positive $k_x$ feels the raised barrier which eventually exactly reverses the direction of the Hall current compared to the simple picture provided in our manuscript. Another important consequence regarding the TAHE conductances' magnitude has also been deduced from that skew reflection

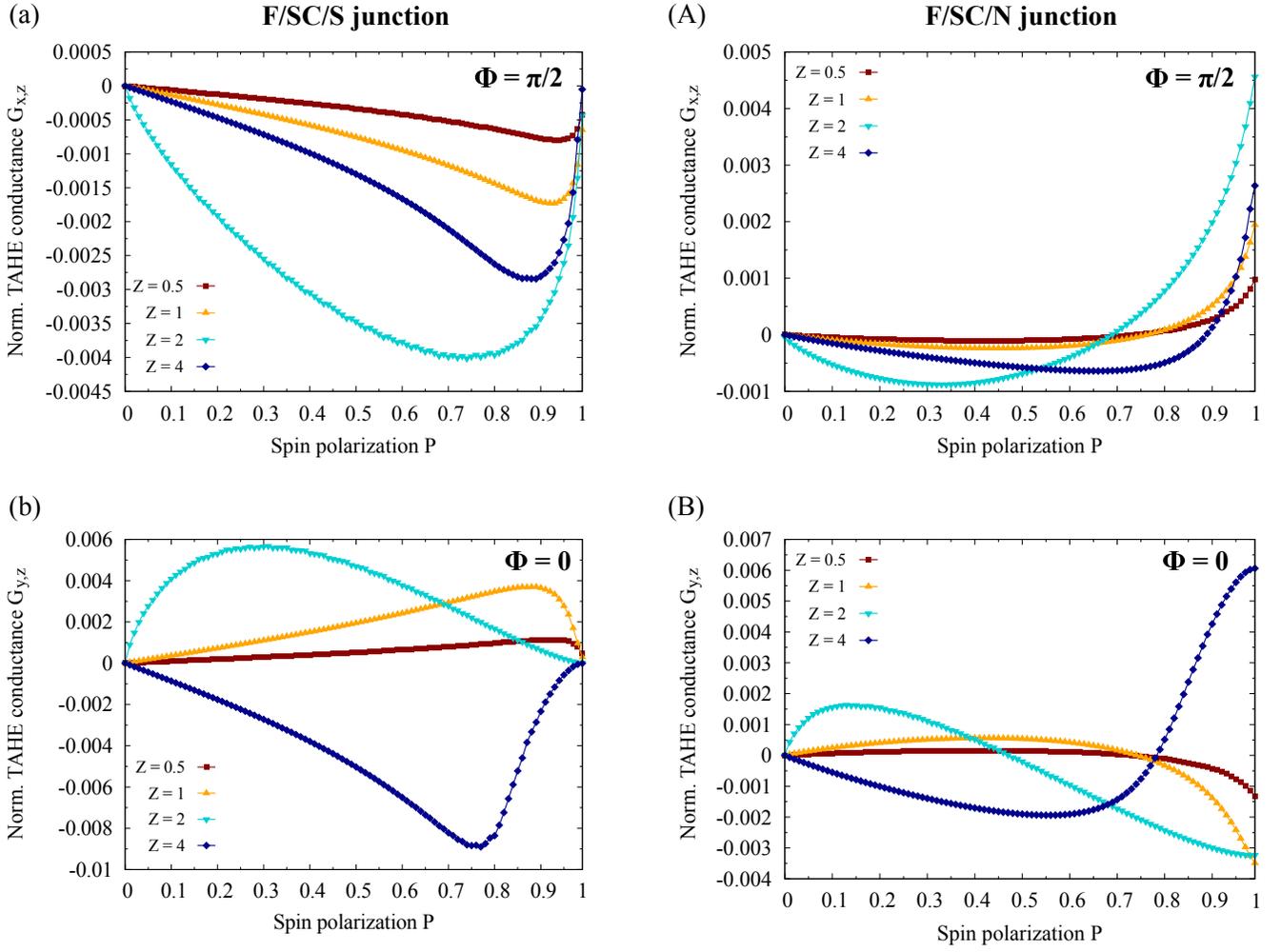

FIG. S6. Calculated dependence of the maximal zero-bias TAHE conductances, (a) $G_{x,z}(\Phi = \pi/2)$ and (b) $G_{y,z}(\Phi = 0)$, normalized to Sharvin's conductance, on the F's spin polarization $P$ ($P = 0.7$ for Fe), and for various indicated barrier strengths $Z$; the SOC parameters are $\alpha = 42.3\,\text{eV\,Å}^2$ and $\beta \approx 19.2\,\text{eV\,Å}^2 \cdot Z$. (A) and (B) show the corresponding normal-state calculations.

picture in the manuscript. Strong barriers always inevitably suppress TAHEs; in the superconducting case even faster than in the normal state. The general symmetry, we observe with respect to $\bar{\lambda}_R$'s sign, is consistent with our symmetry analysis provided in Sec. IV. There, we concluded $G_{x,z}(\Phi) \sim -(\alpha + \beta)\sin\Phi$ and $G_{y,z}(\Phi) \sim (\alpha - \beta)\cos\Phi$; since $\bar{\lambda}_R = (2m\alpha)/\hbar^2 \sim \alpha$ as well as $\bar{\lambda}_D = (2m\beta)/\hbar^2 \sim \beta$, this reduces to

$$G_{x,z}(\Phi = \pi/2) \sim -\bar{\lambda}_R - \bar{\lambda}_D \qquad \text{and} \qquad G_{y,z}(\Phi = 0) \sim \bar{\lambda}_R - \bar{\lambda}_D. \tag{S43}$$

As $\bar{\lambda}_D$ is kept constant within our calculations (specific for the SC layer's material and thickness; see manuscript), $G_{x,z}(\Phi = \pi/2)$ directly merges into $G_{y,z}(\Phi = 0)$ when switching $\bar{\lambda}_R$'s sign, just as our numerical simulations indicate.

Secondly, we present the modulation of the maximal (zero-bias) TAHE conductances with the barrier measure $Z$ in Fig. S5. For the Rashba SOC, we select two representative values of $\alpha = 42.3\,\text{eV\,Å}^2$ and $\alpha = -17.4\,\text{eV\,Å}^2$, respectively; both values lie well within experimentally accessible regimes [S17, S18]. As explained within our qualitative skew reflection model in the preceding paragraph, increasing $Z$ can already be sufficient (depending on the Rashba SOC) to reverse the Hall currents' orientations. Particularly for the two regarded values of Rashba SOC, this applies to $G_{x,z}$ for $\alpha = -17.4\,\text{eV\,Å}^2$ and to $G_{y,z}$ for $\alpha = 42.3\,\text{eV\,Å}^2$; the detailed SOC-dependence has been investigated above. Regarding the TAHE conductances' amplitudes, all previous arguments still hold. The effect is much more pronounced in the superconducting than in the normal-conducting regime at moderate $Z$, but gets enormously damped in the tunneling limit referring to large $Z$ (again even faster in the presence of superconductivity).

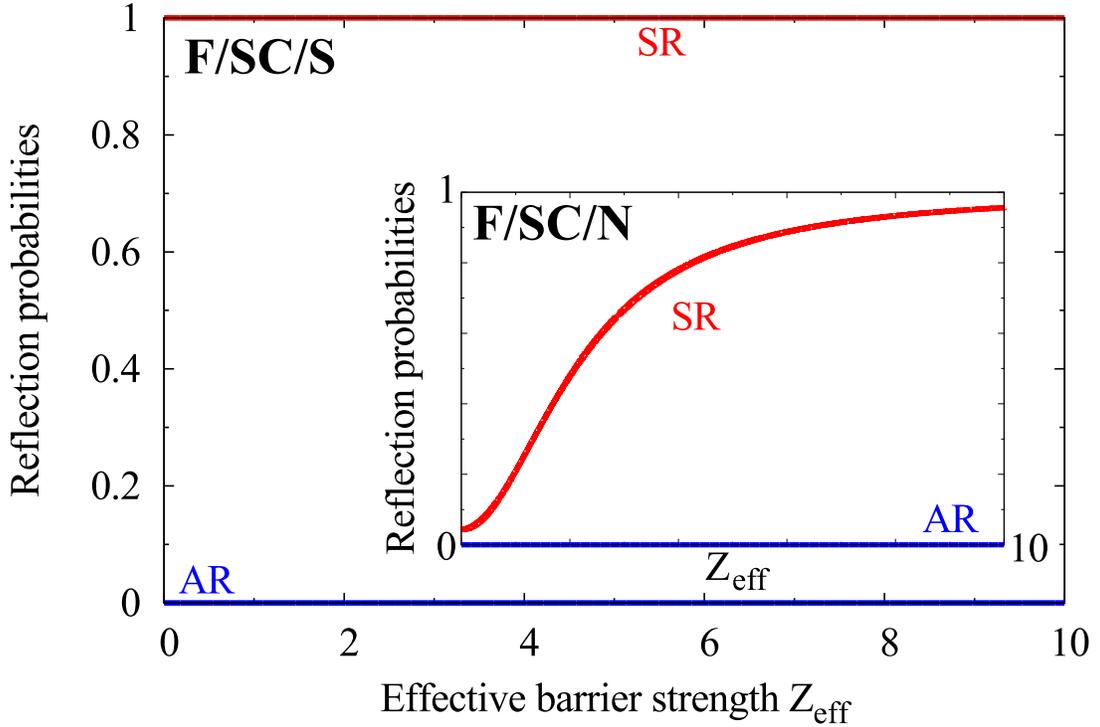

FIG. S7. Calculated (zero-bias) AR and SR probabilities for an incoming spin up electron at the SC interface as a function of the effective barrier parameter $Z_{eff}$ within our simple qualitative framework introduced in the manuscript; the F is *half-metallic* ($P = 1.0$). Since quasiparticle transmissions are forbidden inside the S gap and AR is not possible as there are no minority spin partners, all incoming electrons undergo SR. Therefore, skew reflection at effectively spin- and momentum-dependent barriers cannot produce any spatial charge asymmetry and the Hall currents must vanish. In the normal-conducting state (inset), there are also allowed transmission processes (not explicitly shown) which give rise to the characteristic variations of the SR probability with increasing $Z_{eff}$, necessary for efficient skew SR and to generate measurable Hall currents.

Another indirect evidence for the striking interplay between skew AR and SR can be extracted by studying the impact of the F's spin polarization, $P$, on the TAHE conductances'; so far, we always considered the special case of an iron electrode ($P = 0.7$). Panels (a) and (b) of Fig. S6 illustrate the maximal (zero-bias) TAHE conductances' dependence on $P$ for different $Z$'s and constant Rashba SOC, $\alpha = 42.3 \, \text{eV} \, \text{Å}^2$. For comparison, panels (A) and (B) reveal analogous normal-state calculations. If the junctions' left electrode gets replaced by just a N ($P = 0$), there will no longer be a TAHE as the skew AR and SR contributions of incident up and down spin electrons will exactly compensate. Nevertheless, it is important to observe that even very weak spin polarizations are already sufficient in the superconducting junctions to generate sizable TAHE conductances; our chosen value of $P = 0.7$ for Fe lies close to the optimal value to generate maximal Hall currents. Of particular interest in upcoming experimental studies might be the case of a half-metallic F ($P = 1.0$), which reveals in Fig. S6 clearly distinct features in the superconducting and normal-conducting state. Let us initially understand the S scenario. Since there are only majority spins available close to the Fermi level, majority spins cannot find minority spin partners to undergo (spin-conserving) AR. Within our qualitative picture, all incident electrons must then be specularly reflected (quasiparticle transmissions are not possible inside the S gap), fully independent of the actual effective interfacial scattering potential; see Fig. S7. Just as concluded earlier for the special case of $eV = |\Delta_S|$, this means that incoming electrons no longer experience a skew reflection asymmetry and the TAHE conductance must vanish. However, the spin-flip skew AR and SR conductance contributions, which were not included into the simplified qualitative picture as they have been shown to be negligibly small for Fe-based junctions, play a more important role in the half-metallic junctions so that finally, a small, but finite, TAHE conductance remains. Particularly the presence of (weak) spin-flip AR suggests that still a sizable (mostly) spin-polarized supercurrent response (due to triplet pairing) in the S may be generated, especially interesting for superconducting spintronics. Contrarily in the normal-conducting case, transmission processes need to be considered from the beginning on, even at zero bias. The resulting effective interplay between SR and the transmission processes gives again rise to the characteristic skew SR, which eventually lets pronounced TAHEs emerge—now in the normal-conducting regime and no longer in the superconducting scenario. In summary, the strongly reversing trends of the TAHE conductances' amplitudes in half-metallic systems might provide another important experimental fingerprint to unravel whether the junction is indeed in its purely superconducting phase.

## VIII. ESTIMATION OF TYPICAL TAHE VOLTAGES

For experimental purposes, it may be more convenient to directly measure the TAHE voltage drop instead of the associated conductances. In a F electrode with cubic symmetry and the same size along $\hat{x}$ and $\hat{y}$, the current and voltage components are related as

$$\begin{bmatrix} J_x \\ J_y \\ J_z \end{bmatrix} = \begin{bmatrix} G_\perp & G_{x,y} & G_{x,z} \\ G_{y,x} & G_\perp & G_{y,z} \\ G_{z,x} & G_{z,y} & G_{z,z} \end{bmatrix} \cdot \begin{bmatrix} V_x \\ V_y \\ V_z \end{bmatrix}, \tag{S44}$$

or inversely,

$$\begin{bmatrix} V_x \\ V_y \\ V_z \end{bmatrix} = \begin{bmatrix} R_\perp & R_{x,y} & R_{x,z} \\ R_{y,x} & R_\perp & R_{y,z} \\ R_{z,x} & R_{z,y} & R_{z,z} \end{bmatrix} \cdot \begin{bmatrix} J_x \\ J_y \\ J_z \end{bmatrix}. \tag{S45}$$

Under open circuit conditions, nonzero tunneling current flows along the junction, $J_z \neq 0$, while $J_x = J_y = 0$. Therefore, the voltages are simply given by

$$V_x = R_{x,z} J_z, \tag{S46}$$
$$V_y = R_{y,z} J_z, \tag{S47}$$

as well as

$$V_z = R_{z,z} J_z, \tag{S48}$$

with the TAHE resistances

$$R_{x,z} = \frac{-G_{x,z} G_\perp + G_{x,y} G_{y,z}}{G_{x,z} \left(-G_\perp G_{z,x} + G_{y,x} G_{z,y}\right) + G_{x,y} \left(G_{y,z} G_{z,x} - G_{y,x} G_{z,z}\right) + G_\perp \left(-G_{y,z} G_{z,y} + G_\perp G_{z,z}\right)} \tag{S49}$$

and

$$R_{y,z} = \frac{G_{x,z} G_{y,x} - G_{y,z} G_\perp}{G_{x,z} \left(-G_\perp G_{z,x} + G_{y,x} G_{z,y}\right) + G_{x,y} \left(G_{y,z} G_{z,x} - G_{y,x} G_{z,z}\right) + G_\perp \left(-G_{y,z} G_{z,y} + G_\perp G_{z,z}\right)}. \tag{S50}$$

Accordingly, the tunneling resistance is

$$R_{z,z} = \frac{-G_{x,y} G_{y,x} + G_\perp^2}{G_{x,z} \left(-G_\perp G_{z,x} + G_{y,x} G_{z,y}\right) + G_{x,y} \left(G_{y,z} G_{z,x} - G_{y,x} G_{z,z}\right) + G_\perp \left(-G_{y,z} G_{z,y} + G_\perp G_{z,z}\right)}. \tag{S51}$$

The conductance tensor's "transverse" diagonal components, $G_\perp (= G_{x,x} = G_{y,y})$ describe the F region's response to an applied transverse voltage (without involving quantum mechanical tunneling). Specifically for an iron electrode with length $\ell = 10\,\mu$m and interfacial cross-section $A = \ell^2$, we estimate its magnitude as $G_\perp = \sigma_{Fe} \frac{A}{\ell} \approx 100\,\Omega^{-1}$ [$\sigma_{Fe} \sim 1.0 \times 10^7\,(\Omega m)^{-1}$ is the conductivity of iron [S22]]. In a similar way, $G_{x,y}$ (accounting for the conventional anomalous Hall effect) is expected [S23] to be just about $G_{x,y} \approx 0.8\,\Omega^{-1} \ll G_\perp$ (and for simplicity, we assume $G_{y,x} = G_{x,y}$), which can be neglected in good approximation. To compute typical values for the TAHE voltages, we use the numerically evaluated conductances for the junction with barrier strength $Z = 1$ and applied bias voltage $eV = 0.1|\Delta_S|$; see orange curves in Figs. S2(a) and (b). For the used parameters, the tunneling current is about $J_z = V_z/R_{z,z} = V/R_{z,z} \approx 42$ mA; for the resulting TAHE voltages, we finally obtain $V_x \approx 12\,\mu$V and $V_y \approx -30\,\mu$V, which are indeed sizable TAHE voltages when compared to purely normal-conducting systems [S13, S24].



\end{document}